\documentclass[10pt]{article}
\usepackage{multicol}
\usepackage{graphicx}
\usepackage{amsmath}
\usepackage[a4paper]{geometry}
\usepackage{rotating}
\usepackage{verbatim}
\usepackage{hyperref}

\begin{document}

\begin{center}
{\large\bf Dependence of related parameters on centrality and mass
in a new treatment for transverse momentum spectra in high energy
collisions}

\vskip.75cm

Pei-Pin~Yang$^{1,2,}${\footnote{E-mail: peipinyangshanxi@163.com;
yangpeipin@qq.com}}, Mai-Ying~Duan$^{1,2,}${\footnote{E-mail:
duanmaiying@sxu.edu.cn}},
Fu-Hu~Liu$^{1,2,}${\footnote{Corresponding author. E-mail:
fuhuliu@163.com; fuhuliu@sxu.edu.cn}}

\vskip.25cm

{\small\it $^1$Institute of Theoretical Physics \& Department of
Physics \& State Key Laboratory of Quantum Optics and \\ Quantum
Optics Devices, Shanxi University, Taiyuan, Shanxi 030006,
People's Republic of China

$^2$Collaborative Innovation Center of Extreme Optics, Shanxi
University, \\ Taiyuan, Shanxi 030006, People's Republic of China}

\end{center}

\vskip.5cm

\noindent {\bf Abstract:} We collected the experimental data of
transverse momentum spectra of identified particles produced in
proton-proton ($p$-$p$), deuteron-gold ($d$-Au or $d$-$A$),
gold-gold (Au-Au or $A$-$A$), proton-lead ($p$-Pb or $p$-$A$), and
lead-lead (Pb-Pb or $A$-$A$) collisions measured by the ALICE,
CMS, LHCb, NA49, NA61/SHINE, PHENIX, and STAR collaborations at
different center-of mass energies. The multisource thermal model
at the quark level or the participant quark model is used to
describe the experimental data. The free parameters, the effective
temperature $T$, entropy index-related $n$, and revised index
$a_{0}$, in the revised Tsallis--Pareto-type function are
extracted at the quark level. In most cases, $T$ and $n$ in
central collisions are larger than those in peripheral collisions,
and $a_0$ does not change in different centrality classes. With
the increase in the mass of produced particle or participant
quark, $T$ and $a_0$ increase, and $n$ does not change
significantly. The behaviors of related parameters from $p$-$p$,
$p(d)$-$A$, and $A$-$A$ collisions are similar.

\vskip1.5cm

\begin{multicols}{2}

{\section{Introduction}}

It is believed that quark-gluon plasma (QGP) is possibly generated
in the environment of high temperature and high density formed
within a few microseconds after the little Big Bang of high energy
nucleus-nucleus ($A$-$A$) collisions~\cite{1,2,3,4}. Then the
collision system undergoes likely hydrodynamic evolution and
finally many particles appear after the evolution. In the
evolution, there are at least three main stages: initial, chemical
freeze-out, and kinetic freeze-out. Different types of particles
may experience different stages of evolution and then show
different behaviors of distribution. In addition, different
particles come from similar high energy collision process may have
some similarities in their
distributions~\cite{4a,4b,4c,4d,4e,4f,4g,4h}.

Quantum chromodynamics (QCD) is the fundamental theory of strong
interactions in which quark and gluon are basic free particles.
When the energy for exchanging gluons is very low, the confinement
and chiral symmetry are restored. As a result, quark and gluon are
almost massless. When the energy for exchanging gluons is very
high, the confinement and chiral symmetry are destroyed, which
gives quark rest mass. The characteristics of QGP are topics that
people are more enthusiastic about~\cite{1,2,3,4}. Recent studies
have shown that QGP may be also formed in high-multiplicity events
in high energy proton-proton ($p$-$p$) collisions~\cite{5}.
Studying the transverse momentum spectra of various particles is
one of the promising methods for learning the process of high
energy collisions, and thus the nature of QGP can be understood.
Studying the spectra in a wide range and with different
centralities is an important method to further understand the
mechanisms of QGP hadronization and particle production.

As we know, the calculation of hadron spectrum includes at least
the quark potential model~\cite{6,7} and the lattice
QCD~\cite{8,9,10,11,12}. The former calculates the mass, width,
and other properties of hadrons by constructing the interaction
potential between quarks, and the latter studies the properties of
hadrons mainly from the first principles. For the research of
hadron's transverse momentum spectrum, people mainly study at the
hadron level, except for a few at the quark level. In addition,
the description of the larger transverse momentum (the transverse
momentum being greater than 5 GeV/$c$) is generally described by
the multi-component distribution function. At least, the
two-component distribution function is needed for wide spectra.

In the two-component function, the first component describes the
spectrum in low transverse momentum region, which reflects the
contribution of soft excitation process. The second component
describes the spectrum in medium and high transverse momentum
region, which reflects the contribution of hard scattering
process. Due to the similarities in high energy
collisions~\cite{4a,4b,4c,4d,4e,4f,4g,4h}, the soft and hard
process may show similar characteristics such as the similar
participant quarks. According to our studies~\cite{13,13a,13b} on
the two-component function~\cite{13b}, we may think that the soft
process is performed by the sea quarks and the hard process is
performed by the valence quarks. It is possible that the
contributions of soft and hard processes can be described by a
unified function which reflects the contribution of participant
quarks.

As a practicable method, we hope to use a simple function to
systemize the wide spectra of various particles. In fact, in our
recent work~\cite{13}, in the framework of multisource thermal
model at the quark level~\cite{13a,13b} or participant quark model
and based on the Tsallis statistics~\cite{13c,13d,13e,13f,13ff},
we have used the convolution of two or three revised
Tsallis--Pareto-type (TP-like) functions~\cite{13g,13h,13i,13j} to
fit the available spectra of various particles produced in $p$-$p$
collisions at 200 GeV at the Relativistic Heavy Ion Collider
(RHIC) and at 2.76 and 13 TeV at the Large Hadron Collider (LHC).
The part behaviors of the considered convolution function are
studied due to changing the main parameters which includes the
effective temperature $T$, entropy index-related $n$, and revised
index $a_0$.

To see the behaviors of main parameters in the considered
convolution function, we study the dependence of $T$, $n$, and
$a_0$ on the event centrality $C$, the rest mass $m_0$ of produced
particles, and the constituent mass $m_q$ (\(q=u\), \(d\), \(s\),
\(c\), or \(b\)) of participant quarks in this paper. To compare
conveniently, we collect i) the data on $p$-$p$ collisions at
2.76, 5.02, and 7 TeV measured by the ALICE~\cite{14,15},
CMS~\cite{16}, and LHCb~\cite{17} Collaborations, respectively,
ii) the data on deuteron-gold ($d$-Au or $d$-$A$) collisions at
200 GeV measured by the PHENIX collaboration~\cite{18,19,20,21},
iii) the data on gold-gold (Au-Au or $A$-$A$) collisions at 200
GeV measured by the PHENIX~\cite{18,22,20} and STAR~\cite{23,24}
Collaborations, iv) the data on proton-lead ($p$-Pb or $p$-$A$)
collisions at 5.02 and 8.16 TeV measured by the ALICE
collaboration~\cite{25,26,27,28}, and v) the data on lead-lead
(Pb-Pb or $A$-$A$) collisions at 2.76 and 5.02 TeV measured by the
ALICE collaboration~\cite{29,15,30,31}. To test further, we also
compare the convolution function with the data from $p$-$p$,
Au-Au, and Pb-Pb collisions over a low energy range from 6.3 to 39
GeV.

The remainder of this paper is structured as follows. The
formalism and method are shortly described in Section 2. Results
and discussion are given in Section 3. In Section 4, we summarize
our main observations and conclusions.
\\

{\section{Formalism and method}}

Hadrons are composed of more basic components such as quarks.
According to the multisource thermal model at the quark
level~\cite{13a,13b} or the participant quark model discussed in
our recent work~\cite{13}, we first assume that participant quark
is isotropic in its rest frame, and hadron is contributed by
participant quarks according to the combination. In the
multisource thermal model, it is assumed that one, two, or more
sources emit particles due to different generation mechanisms,
source temperatures, and event samples. In a given event sample,
particles with the same source temperature are generated by the
same mechanism and are considered to be emitted from the same
source. Participant quarks are considered to be the contributor
quarks in the participant quark model.

In this article, we want to study the transverse momentum spectra
with different centralities and larger transverse momentum range
at the quark level. In our previous work~\cite{13}, based on the
Tsallis statistics~\cite{13c,13d,13e,13f,13ff}, we revised the
Tsallis--Pareto-type (TP-type) function~\cite{13g,13h,13i,13j} to
be the TP-like function. That is, we assume that the transverse
momentum (\(p_{\rm T}\)) spectrum of hadrons with rest mass
\(m_{0}\) satisfies the TP-like function,
\begin{align}
f(p_{\rm T})= C_0p_{\rm T}^{a_0}\left( 1+ \frac{\sqrt{p_{\rm
T}^2+m_0^2}-m_0}{nT} \right)^{-n}.
\end{align}
Here the effective temperature \(T\), entropy index-related \(n\),
and revised index \(a_0\) are free parameters, and \(C_0\) is the
normalization constant related to free parameters. It should be
noted that \(a_0\) is a real number with non-dimension. Although
the introduction of \(a_0\) is a technical treatment only, it
determines the bending degree of curve in low $p_{\rm T}$ region
and then affects the slope of curve in medium and high $p_{\rm T}$
region due to the limitation of normalization. We have the
dimensions of \(p_{\rm T}^{a_0}\) and \(C_0\) to be \(({\rm
GeV}/c)^{a_0}\) and \(({\rm GeV}/c)^{-{a_0}-1}\), respectively.
Thus, the dimension of \(C_0p_{\rm T}^{a_0}\) is \(({\rm
GeV}/c)^{-1}\) which is the same as the fit function.

In the combined quarks that form the hadron, the contribution of
the $i$-th participant quark to the transverse momentum of hadron
complies with
\begin{align}
f_i(p_{ti})= C_ip_{ti}^{a_0} \left( 1+
\frac{\sqrt{p_{ti}^2+m_{0i}^2}-m_{0i}}{nT} \right)^{-n}.
\end{align}
Here $f_i(p_{ti})$ is the probability density function of the
transverse momentum $p_{ti}$ contributed by the $i$-th participant
quark, $m_{0i}$ is the constituent mass of the $i$-th participant
quark, and $C_{i}$ is the normalization constant. A meson is a
combination of two quarks, so $i=1$ or 2 for mesons. A baryon is a
combination of three quarks, so $i=1$, 2, or 3 for baryons. For
tetraquark state, we have \(i=1\), 2, 3, or 4. For pentaquark
state, we have \(i=1\), 2, 3, 4, or 5. For six-quark state, we
have \(i=1\), 2, 3, 4, 5, or 6. Equations (1) and (2) are
essentially TP-like functions, but $m_{0}$ in Eq. (1) refers to
the rest mass of the hadron described, and $m_{0i}$ in Eq. (2)
refers to the constituent mass of the $i$-th participant quark. If
the $i$-th participant quark is \(q\), we have \(m_{0i}=m_q\).

Transverse momentum spectrum of mesons is the convolution of two
TP-like functions, We have
\begin{align}
f(p_{\rm T})&=\int_0^{p_{\rm T}} f_1(p_{t1})f_2(p_{\rm T}-p_{t1})dp_{t1} \nonumber\\
&=\int_0^{p_{\rm T}} f_2(p_{t2})f_1(p_{\rm T}-p_{t2})dp_{t2}.
\end{align}
Here $f_1(p_{t1})$ and $f_2(p_{t2})$ are the probability density
functions of transverse momenta contributed by the first and
second participant quarks, respectively, and $f(p_{\rm T})$ is the
probability density function for $p_{\rm T}$ of mesons.

The transverse momentum spectrum of baryons is the convolution of
three TP-like functions. The convolution of the first two TP-like
function is:
\begin{align}
f_{12}(p_{t12})&=\int_0^{p_{t12}} f_1(p_{t1}) f_2(p_{t12}-p_{t1})
dp_{t1} \nonumber\\
&=\int_0^{p_{t12}} f_2(p_{t2}) f_1(p_{t12}-p_{t2}) dp_{t2}.
\end{align}
Here $f_{12}(p_{t12})$ is the probability density function of
transverse momentum contributed by the first two participant
quarks. After the first two TP-like functions are convoluted, they
are convoluted with the third TP-like function:
\begin{align}
f(p_{\rm T})&=\int_0^{p_{\rm T}} f_{12}(p_{t12}) f_3(p_{\rm
T}-p_{t12}) dp_{t12}
\nonumber\\
&=\int_0^{p_{\rm T}} f_3(p_{t3}) f_{12}(p_{\rm T}-p_{t3}) dp_{t3}.
\end{align}
Here $f(p_{\rm T})$ is the probability density function for
$p_{\rm T}$ of baryons.

Equation (1) is used to describe the hadron spectra at the hadron
level. Equations (3) and (5) are used to describe the meson and
baryon spectra at the quark level, respectively. Although Eq. (1)
is approximately successful in the fits of transverse momentum
spectra of various particles in our recent work~\cite{13}, we use
Eqs. (3) and (5) in this paper to fit more accurately the
transverse momentum spectra of hadrons at the quark level. In Eqs.
(3)--(5), no matter how many participant quarks are considered,
the number of parameters is always four which includes three free
parameters and a normalization constant. That is to say that, we
use the same parameters for different participant quarks in a
given hadron.
\\

{\section{Results and discussion}}

{\subsection{Comparison with data by Eqs. (3) and (5)}}

Figure 1(a) shows the $p_{\rm T}$ spectra, $(1/2\pi p_{\rm
T})d^2N/dp_{\rm T}dy$, of different hadrons such as
$\pi^+$+$\pi^-$, $K^+$+$K^-$, and $p$+$\bar p$ with pseudorapidity
$|\eta|<0.8$ and $\phi$ with rapidity $|y|<0.5$ produced in
inelastic (INEL) $p$-$p$ collisions at 2.76 TeV, where $N$ denotes
the number of hadrons. The symbols represent the data measured by
the ALICE Collaboration~\cite{14,15} and the curves are our
results fitted by Eq. (3) for mesons or (5) for baryons. Figure
1(b) shows the $p_{\rm T}$ spectra, $d\sigma/dp_{\rm T}$, of
prompt $\psi(2S)$ and $\psi(2S)$ from $b$ with $2<y<4.5$ produced
in INEL $p$-$p$ collisions at 7 TeV and the $p_{\rm T}$ spectra,
$Bd^2\sigma/dp_{\rm T}dy$, of $\Upsilon(1S)$ with $|y|<2.4$
produced in INEL $p$-$p$ collisions at 5.02 TeV, where $\sigma$
denotes the cross-section and $B$ denotes the branch ratio. The
symbols represent the data measured by the LHCb~\cite{17} and
CMS~\cite{16} Collaborations respectively, and the curves are our
results fitted by Eq. (3). Different symbols represent the spectra
of different particles. Some of them are scaled by multiplying
different factors, as marked in the panels. The values of free
parameters ($T$, $n$, and $a_0$), normalization constant ($N_0$ or
$\sigma_0$), $\chi^2$, and number of degree of freedom (ndof)
obtained from Eqs. (3) or (5) are listed in Table 1. The quark
structures are also listed in Table 1. In the present fitting
process and in the following discussion, the constituent masses of
various quarks and the rest masses of various hadrons are cited
from ``The Review of Particle Physics (2020)"~\cite{31a}. One can
see that the fit equation is in agreement with the data measured
by the ALICE, CMS, and LHCb Collaborations at the LHC.

\begin{figure*}[htbp]
\begin{center}
\includegraphics[width=13.0cm]{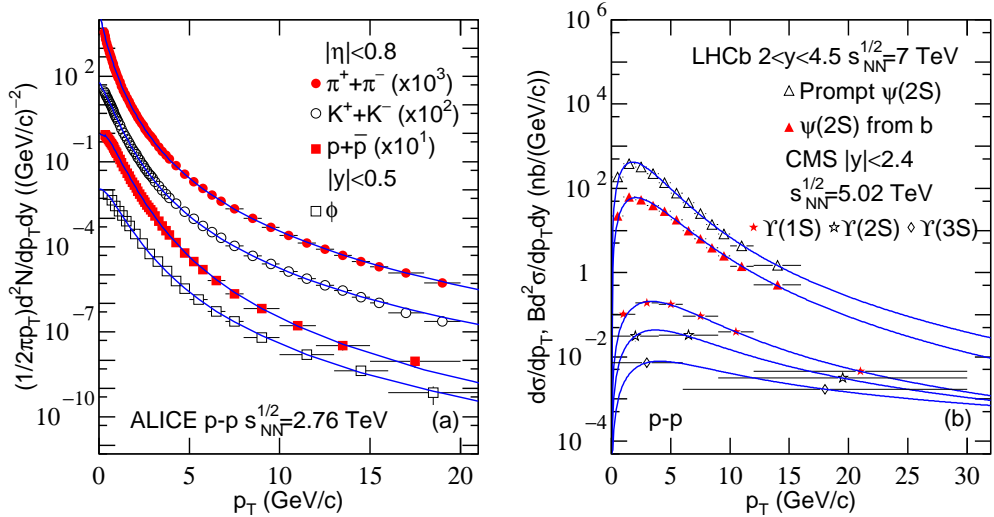}
\end{center}
\caption{\small The transverse momentum spectra of different
hadrons produced in INEL $p$-$p$ collisions at $\sqrt{s_{\rm
NN}}=2.76$, 5.02, and 7 TeV, where $\sqrt{s_{\rm NN}}$ ($s_{\rm
NN}^{1/2}$ in the panels) denotes the center-of-mass energy per
nucleon pair and can be simplified to $\sqrt{s}$ for $p$-$p$
collisions. Different symbols represent the spectra of different
particles measured by the ALICE~\protect\cite{14,15},
CMS~\protect\cite{16}, and LHCb~\protect\cite{17} Collaborations,
where some of them are scaled by different factors marked in the
panels. The curves are our fitted results by using Eqs. (3) for
mesons and (5) for barons.} \vspace{0.5cm}
\end{figure*}

\begin{figure*}[!hb]
\begin{center}
\includegraphics[width=13.0cm]{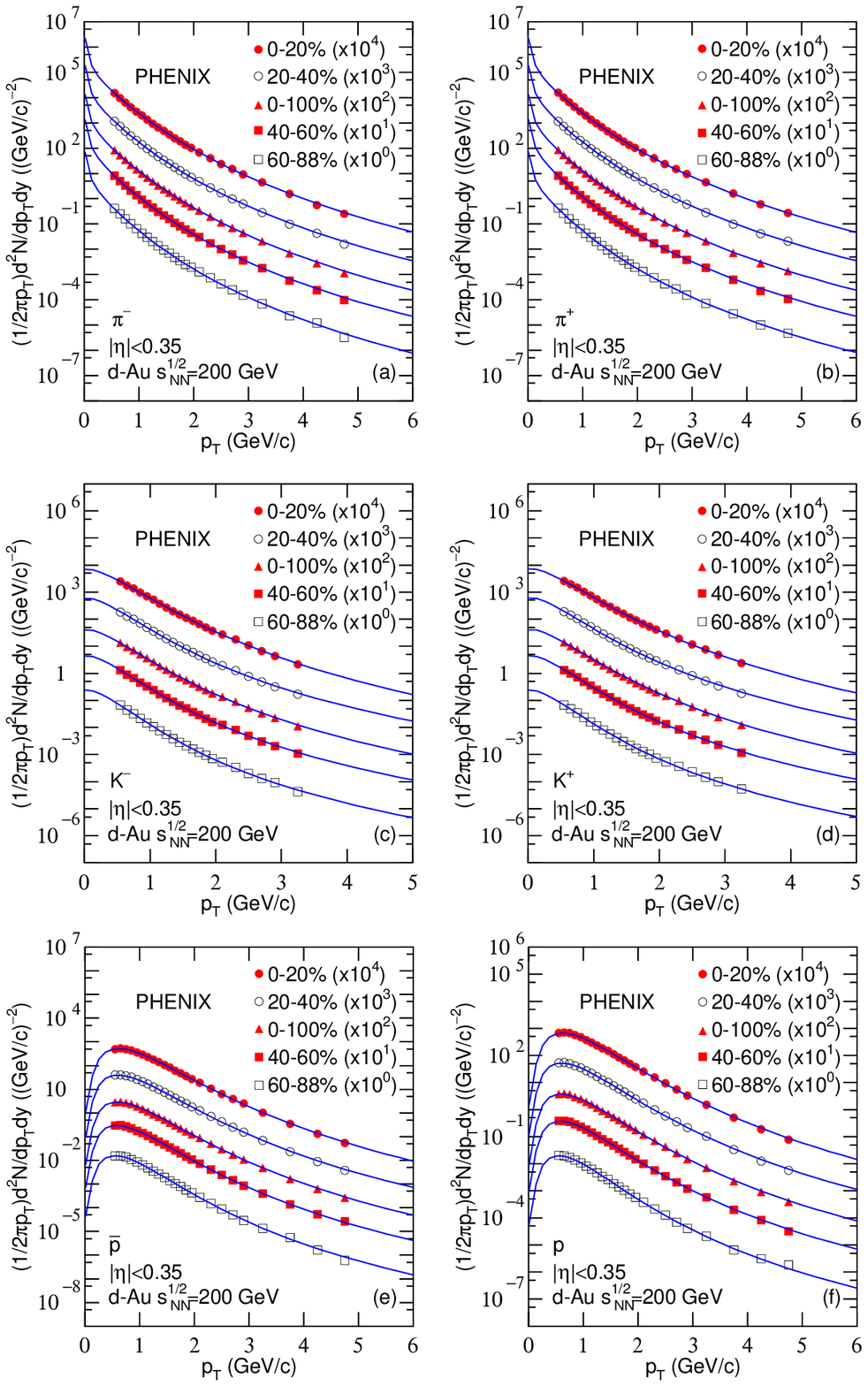}
\end{center}
\end{figure*}

\begin{figure*}[!htb]
\begin{center}
\includegraphics[width=13.0cm]{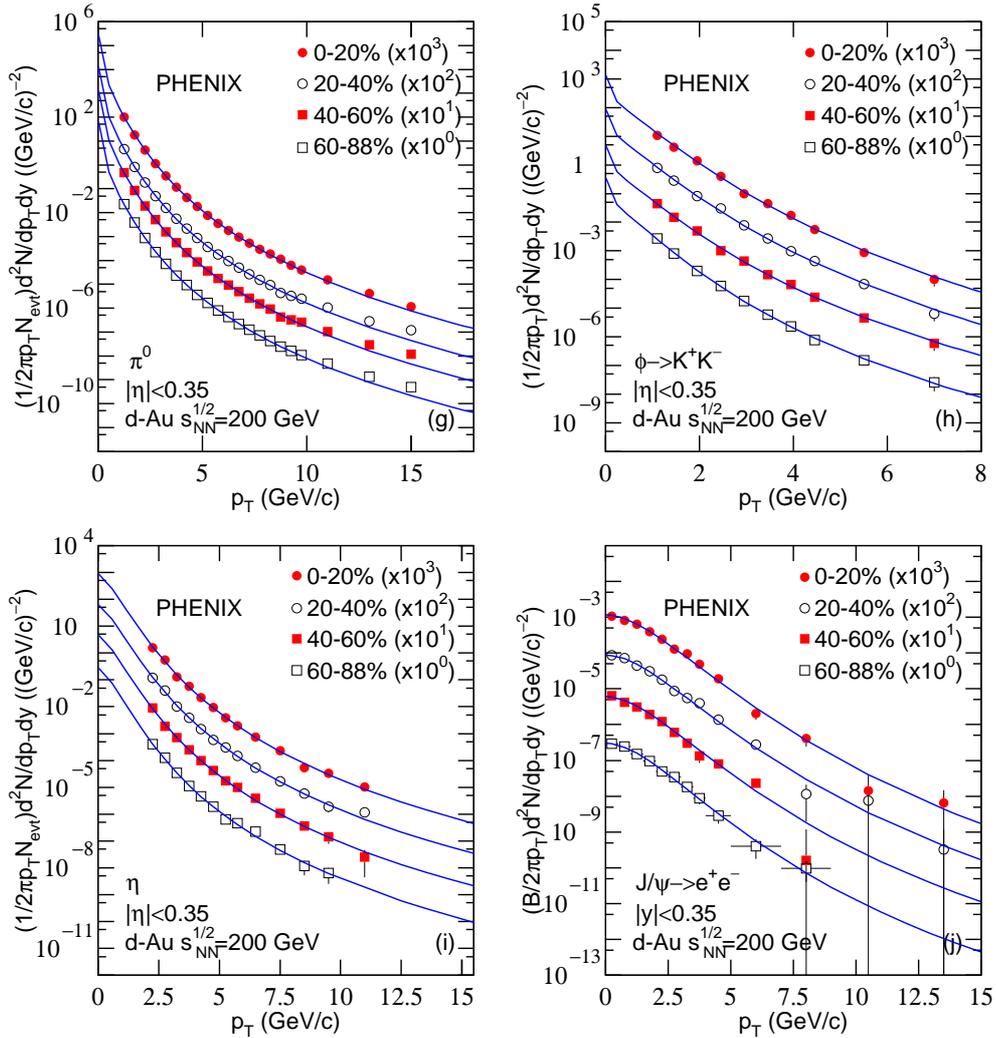}
\caption{\small The transverse momentum spectra of (a) $\pi^-$,
(b) $\pi^+$, (c) $K^-$, (d) $K^+$, (e) $\bar p$, (f) $p$, (g)
$\pi^0$, (h) $\phi\rightarrow K^+K^-$, (i) $\eta$, and (j)
$J/\psi\rightarrow e^+e^-$ with (a)--(i) $|\eta|<0.35$ (or (j)
$|y|<0.35$) in different centrality classes produced in $d$-Au
collisions at 200 GeV. Different symbols represent different
centrality classes measured by the PHENIX
Collaboration~\protect\cite{18,19,20,21}. Some of them are scaled
by different factors marked in the panels. The curves are our
fitted results by using Eqs. (3) for mesons and (5) for baryons.}
\end{center}

\end{figure*}

\end{multicols}
\begin{table*}[!htb]
{\small Table 1. Values of $T$, $n$, $a_0$, $N_{0}$ (or
$\sigma_0$), $\chi^2$, and ndof corresponding to the curves in
Fig. 1, where the normalization constant $\sigma_0$ is for the
cases of prompt $\psi(2S)$, $\psi(2S)$ from $b$, and
$\Upsilon(1S)$ presented in Fig. 1(b). All the free parameters are
extracted at the quark level. \vspace{-0.5cm}
\begin{center}
\begin{tabular} {cccccccccccc}\\ \hline\hline
Particle & $T$ (GeV) & $n$ & $a_0$ & $N_0$ ($\sigma_0$ (nb)) & $\chi^2$/ndof \\
(quark structure) &  &  &  &  & \\
\hline
$\pi^++\pi^-$ ($u\bar d$, $d\bar u$)     & $0.216\pm0.001$ & $5.047\pm0.023$ & $-0.500\pm0.004$ & $4.102\pm0.032$ & 60/59\\
$K^++K^-$ ($u\bar s$, $s\bar u$)         & $0.203\pm0.001$ & $5.136\pm0.023$ & $-0.100\pm0.003$ & $0.483\pm0.003$ & 16/54\\
$p+\bar p$ ($uud$, $\bar u\bar u\bar d$) & $0.205\pm0.001$ & $6.128\pm0.020$ & $-0.130\pm0.002$ & $(5.176\pm0.063)\times10^{-3}$ & 12/45\\
$\phi$ ($s\bar s$)                    & $0.272\pm0.001$ & $5.362\pm0.022$ & $0.000\pm0.002$  & $(4.490\pm0.054)\times10^{-2}$ & 5/17\\
prompt $\psi(2S)$ ($c\bar c$)         & $0.538\pm0.002$ & $4.962\pm0.023$ & $0.300\pm0.003$  & $1439.485\pm6.997$ & 16/8\\
$\psi(2S)$ from $b$ ($c\bar c$)       & $0.675\pm0.001$ & $5.135\pm0.014$ & $0.297\pm0.002$  & $244.885\pm1.999$ & 9/8\\
$\Upsilon(1S)$ ($b\bar b$)            & $0.668\pm0.002$ & $2.918\pm0.021$ & $0.297\pm0.003$  & $6.769\pm0.047$ & 5/2\\
\hline
\end{tabular}%
\end{center}}
\end{table*}
\begin{multicols}{2}

\begin{figure*}[!hb]
\begin{center}
\includegraphics[width=13.0cm]{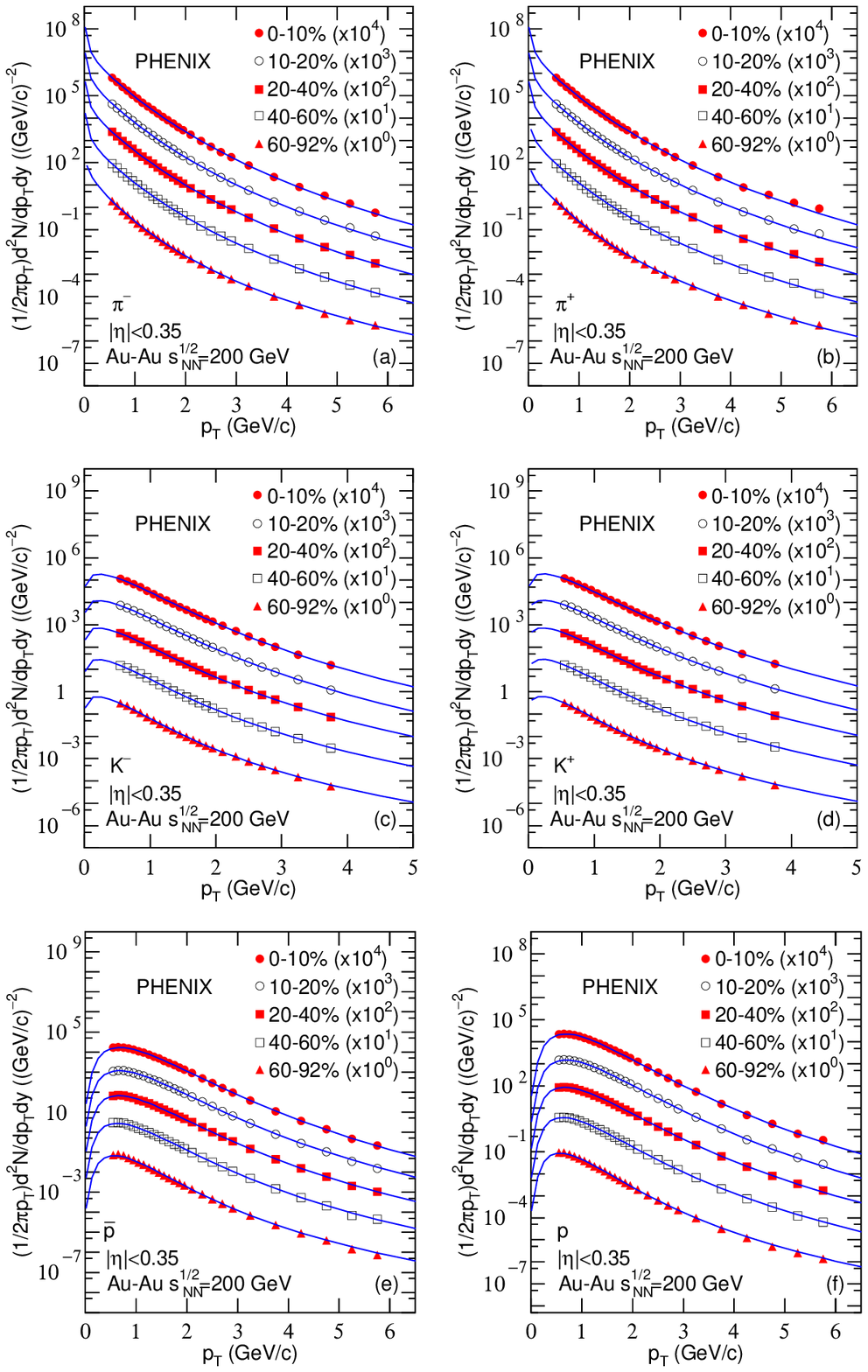}
\end{center}
\end{figure*}

\begin{figure*}[!htb]
\begin{center}
\includegraphics[width=13.0cm]{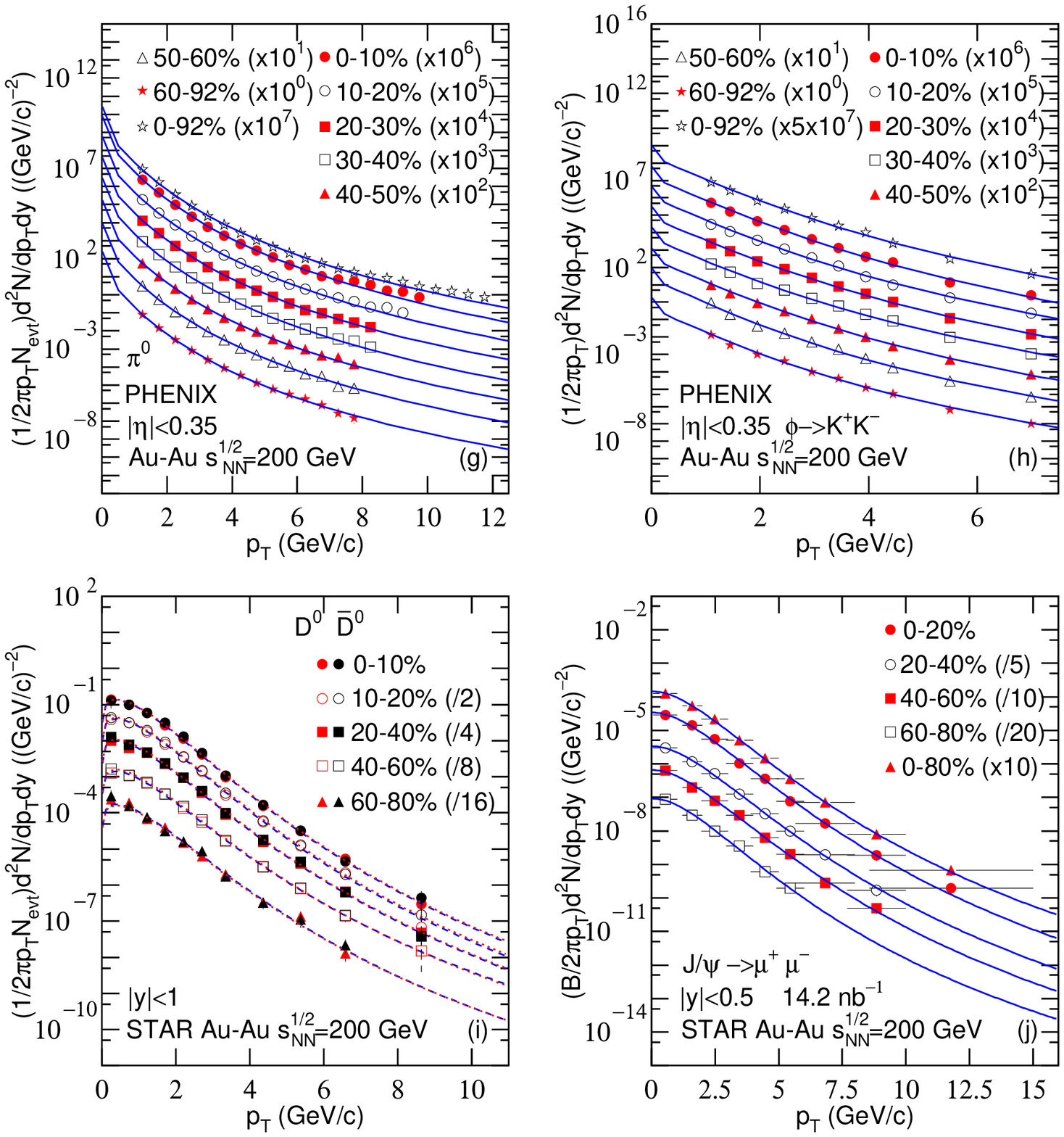}
\end{center}
\caption{\small The transverse momentum spectra of (a) $\pi^-$,
(b) $\pi^+$, (c) $K^-$, (d) $K^+$, (e) $\bar p$, (f) $p$, (g)
$\pi^0$, (h) $\phi\rightarrow K^+K^-$, (i) $D^0$ ($\bar D^0$), and
(j) $J/\psi\rightarrow \mu^+\mu^-$ with mid-$\eta$ or mid-$y$
produced in Au-Au collisions at 200 GeV. Different symbols
represent different centrality classes measured by (a)--(h) the
PHENIX Collaboration~\protect\cite{18,20,22} and (i)--(j) the STAR
Collaboration~\protect\cite{23,24}. Some of them are scaled by
different factors marked in the panel. The curves are our fitted
results by using Eqs. (3) for mesons and (5) for baryons, where
the dotted and dashed curves fit the spectra of $D^0$ and $\bar
D^0$ respectively.}
\end{figure*}

Figure 2 shows the centrality dependent $p_{\rm T}$ spectra of (a)
$\pi^-$, (b) $\pi^+$, (c) $K^-$, (d) $K^+$, (e) $\bar p$, (f) $p$,
(g) $\pi^0$, (h) $\phi\rightarrow K^+K^-$, (i) $\eta$, and (j)
$J/\psi\rightarrow e^+e^-$ with $|\eta|<0.35$ (or $|y|<0.35$)
produced in $d$-Au collisions at 200 GeV. The equation of $p_{\rm
T}$ spectra for panels (a)--(f) and (h) is $(1/2\pi p_{\rm
T})d^2N/dp_{\rm T}dy$, for (g) and (i) is $(1/2\pi p_{\rm T}
N_{\rm evt})d^2N/dp_{\rm T}dy$, and for (j) is $(B/2\pi p_{\rm
T})d^2N/dp_{\rm T}dy$. Here $N_{\rm evt}$ denotes the number of
events. Different symbols represent the experimental data with
different centrality classes in $d$-Au collisions at 200 GeV,
measured by the PHENIX Collaboration~\cite{18,19,20,21}. As shown
in the panels, some of them are scaled by multiplying different
factors. The curves are our fitting results, which are
approximately in agreement with the experimental data of $d$-Au
collisions measured by the PHENIX Collaboration at the RHIC. The
related parameters are listed in Table 2.

Similar with Fig. 2, Figure 3 shows the centrality dependent
$p_{\rm T}$ spectra of (a) $\pi^-$, (b) $\pi^+$, (c) $K^-$, (d)
$K^+$, (e) $\bar p$, (f) $p$, (g) $\pi^0$, (h) $\phi\rightarrow
K^+K^-$, (i) $D^0$ ($\bar D^0$), and (j) $J/\psi\rightarrow
\mu^+\mu^-$ produced in Au-Au collisions at 200 GeV. The equation
of $p_{\rm T}$ spectra for panels (a)--(f) and (h) is $(1/2\pi
p_{\rm T})d^2N/dp_{\rm T}dy$, for (g) and (i) is $(1/2\pi p_{\rm
T} N_{\rm evt})d^2N/dp_{\rm T}dy$, and for (j) is $(B/2\pi p_{\rm
T})d^2N/dp_{\rm T}dy$, The experimental data with given $|\eta|$
and $|y|$ ranges marked in the panels are measured by the
PHENIX~\cite{18,22,20} and STAR~\cite{23,24} Collaborations,
respectively. Our fitting results, as shown by the curves, are
consistent with the experimental data of Au-Au collisions measured
by the PHENIX and STAR Collaborations at the RHIC. See Table 3 for
the fitting parameters.

Similar with Figs. 2 and 3, Figures 4(a)--4(e) show the centrality
dependent $p_{\rm T}$ spectra, (a)--(c) $(1/2\pi p_{\rm
T})d^2N/dp_{\rm T}dy$ and (d) and (e) $(1/N_{\rm evt})d^2N/dp_{\rm
T}dy$, of (a) $\pi^++\pi^-$, (b) $K^++K^-$, (c) $p+\bar p$, (d)
$\Sigma(1385)^+$, and (e) $(\Xi(1530)^0+\bar\Xi(1530)^0)/2$ with
$-0.5<y<0$ measured by the ALICE Collaboration~\cite{25,26} in
$p$-Pb collisions at 5.02 TeV, accompanied by the $p_T$ spectra of
the mentioned particles in non-single diffraction (NSD) $p$-$p$
collisions at the same energy per nucleon pair. Figure 4(f) shows
the $p_{\rm T}$ spectra, $d^2N/dp_{\rm T}dy$, of $D^0$, $D_s^+$,
$D^+$, and $D^{*+}$ with $-0.96<y<0.04$, as well as $\Upsilon(1S)$
with $2.03<y<3.53$ and $-4.46<y<-2.96$, measured by the ALICE
Collaboration~\cite{27,28} in minimum-bias (MB) $p$-Pb collisions
at 5.02 and 8.16 TeV. The curves represent our fitted results and
the related parameters are listed in Table 4. The fitting results
are in agreement with the experimental data of $p$-Pb and $p$-$p$
collisions measured by the ALICE Collaboration at the LHC.

Similar with Figs. 2, 3, and 4(a)--4(e), Figure 5 shows the
transverse momentum spectra, (a)--(e) $(1/2\pi p_{\rm
T})d^2N/dp_{\rm T}dy$, (f) and (g) $dN/dp_{\rm T}$, and (h)
$d^2N/dp_{\rm T}dy$, of (a) $\pi^++\pi^-$, (b) $K^++K^-$, (c)
$p+\bar p$, (d) $(K^{*0}+\bar K^{*0})/2$, (e) $\phi$, (f) $D^+$
($D^0$), (g) $D_s^+$ ($D^{*+}$), and (h) inclusive $J/\psi$
produced in Pb-Pb collisions at 2.76 TeV for panels (a)--(e) and
5.02 TeV for (f)--(h), with different $|\eta|$ or $|y|$ and
centrality classes as marked in the panels. The symbols represent
the experimental data measured by the ALICE
Collaboration~\cite{29,15,30,31}. The curves are our fitted
results and the related parameters are listed in Table 5. The
curves fit well with the experimental data measured by the ALICE
Collaboration on Pb-Pb collisions at the LHC.
\\

\begin{figure*}
\begin{center}
\includegraphics[width=13.0cm]{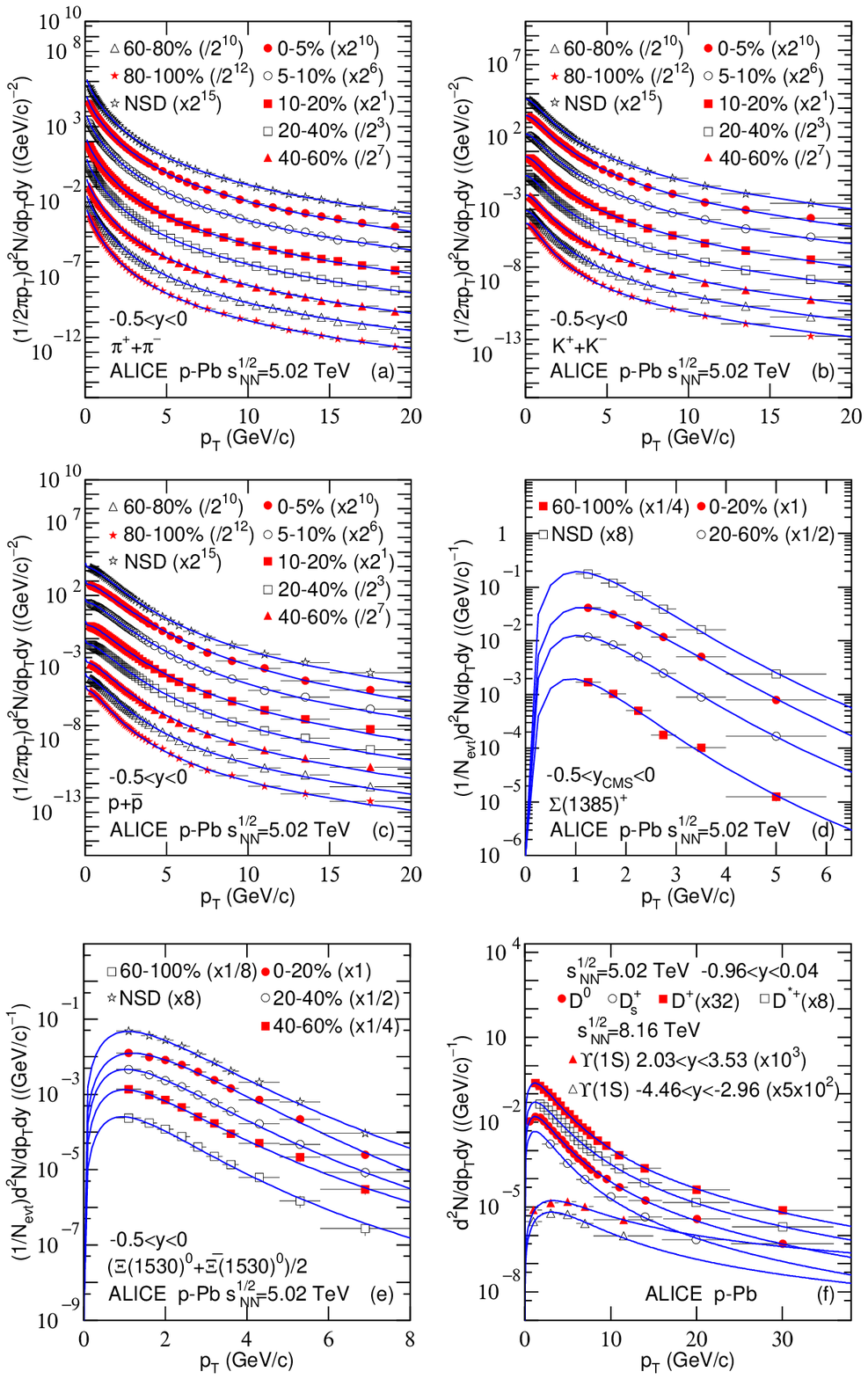}
\end{center}\vspace{-0.5cm}
\caption{\small The transverse momentum spectra of (a)
$\pi^++\pi^-$, (b) $K^++K^-$, (c) $p+\bar p$, (d)
$\Sigma(1385)^+$, and (e) $(\Xi(1530)^0+\bar\Xi(1530)^0)/2$ with
$-0.5<y<0$ in different centrality $p$-Pb and NSD $p$-$p$
collisions at 5.02 TeV, as well as of (f) $D^0$, $D_s^+$, $D^+$,
and $D^{*+}$ with $-0.96<y<0.04$, as well as $\Upsilon(1S)$ with
$2.03<y<3.53$ and $-4.46<y<-2.96$, in MB $p$-Pb collisions at 5.02
and 8.16 TeV. Different symbols represent the experimental data in
different cases measured by the ALICE
Collaboration~\protect\cite{25,26,27,28}. Some of them are scaled
by different factors marked in the panel. The curves are our
fitted results by using Eqs. (3) for mesons and (5) for baryons.}
\end{figure*}

\begin{figure*}[!hb]
\begin{center}
\includegraphics[width=13.0cm]{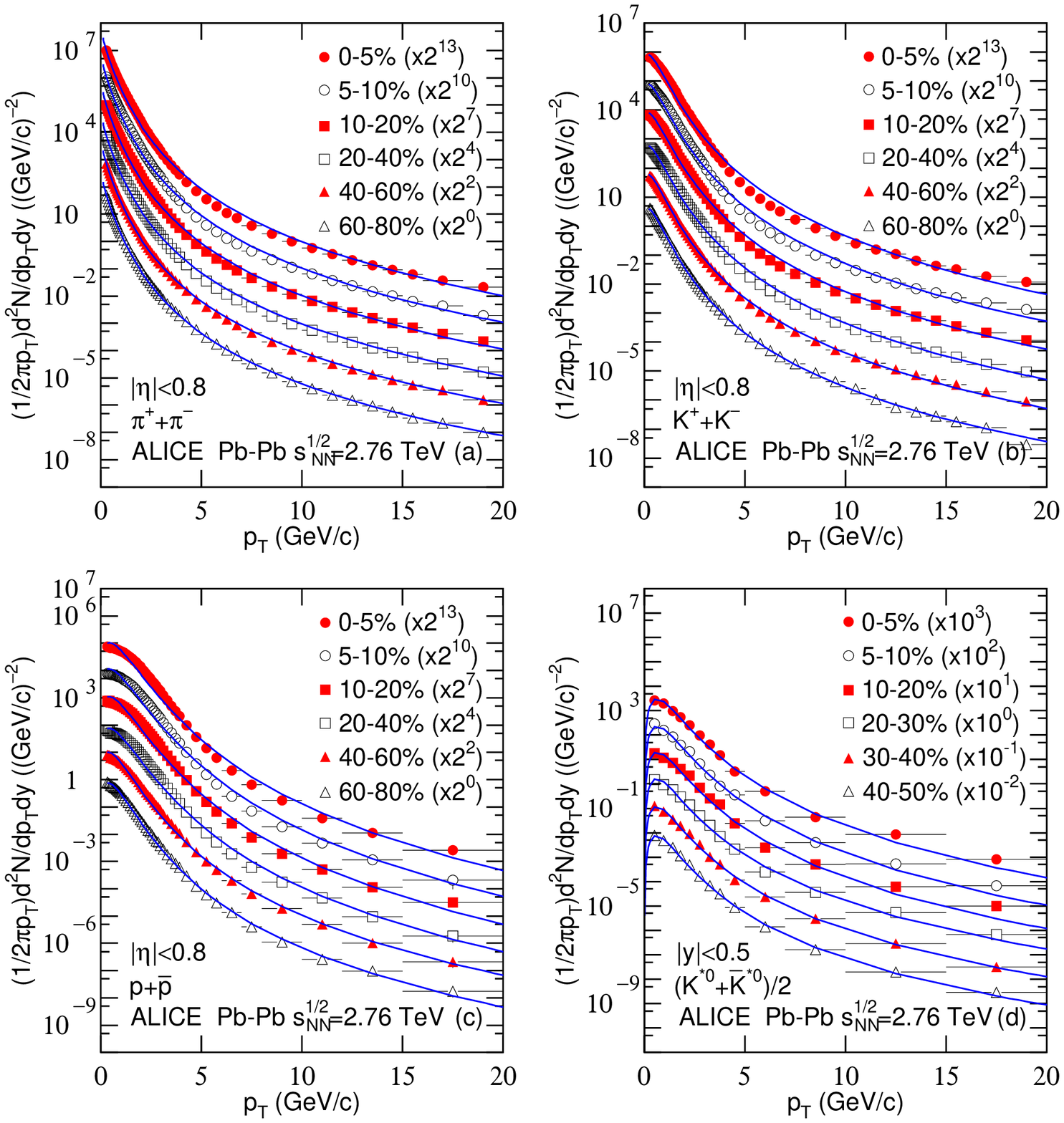}
\end{center}
\end{figure*}

\begin{figure*}[!htb]
\begin{center}
\includegraphics[width=13.0cm]{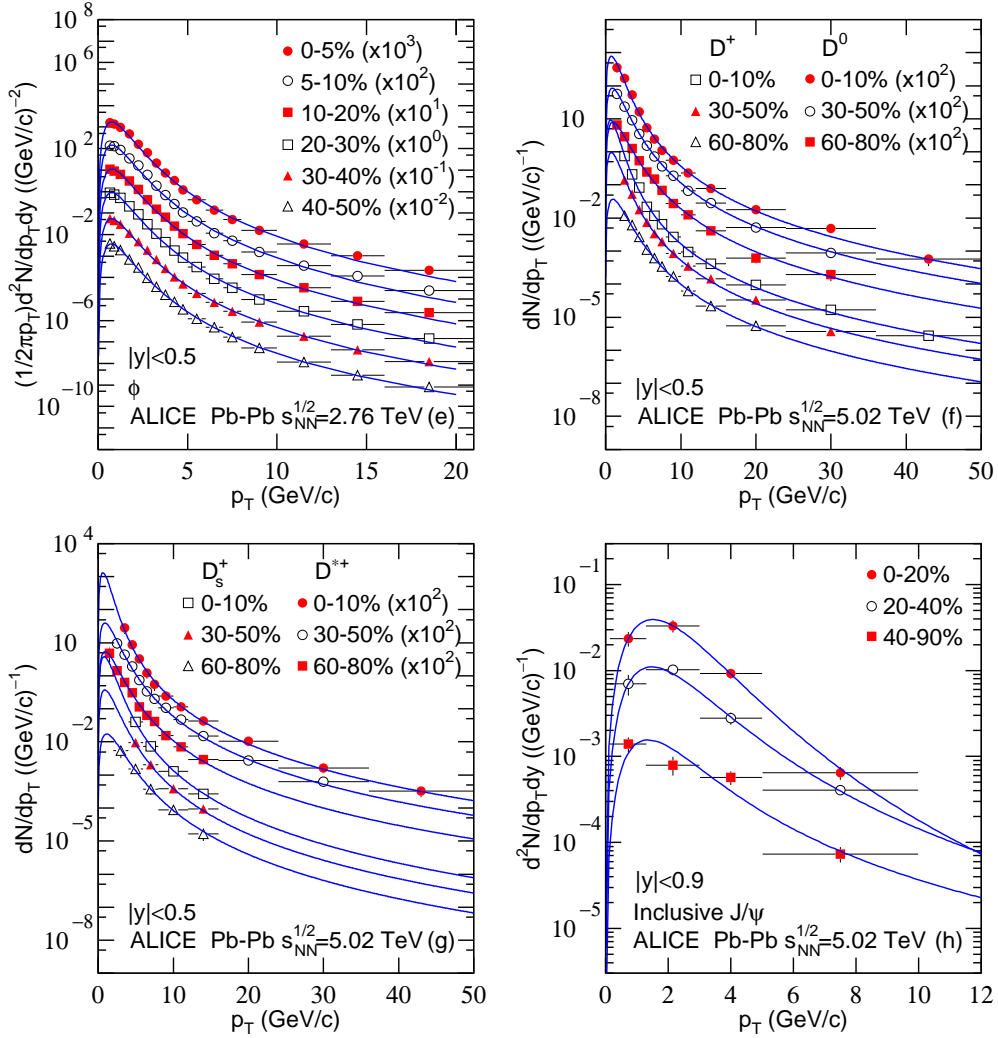}
\end{center}
\caption{\small The transverse momentum spectra of (a)
$\pi^++\pi^-$, (b) $K^++K^-$, (c) $p+\bar p$, (d) $(K^{*0}+\bar
K^{*0})/2$, (e) $\phi$, (f) $D^+$ ($D^0$), (g) $D_s^+$ ($D^{*+}$),
and (h) inclusive $J/\psi$ produced in Pb-Pb collisions at
(a)--(e) 2.76 TeV and (f)--(h) 5.02 TeV with different $|\eta|$ or
$|y|$ and centrality classes. The symbols represent the
experimental data measured by the ALICE
Collaboration~\protect\cite{29,15,30,31}. Some of them are scaled
by different factors. The curves are our fitted results by using
Eqs. (3) for mesons and (5) for baryons.}
\end{figure*}

\begin{figure*}[!htb]
\begin{center}
\includegraphics[width=13.0cm]{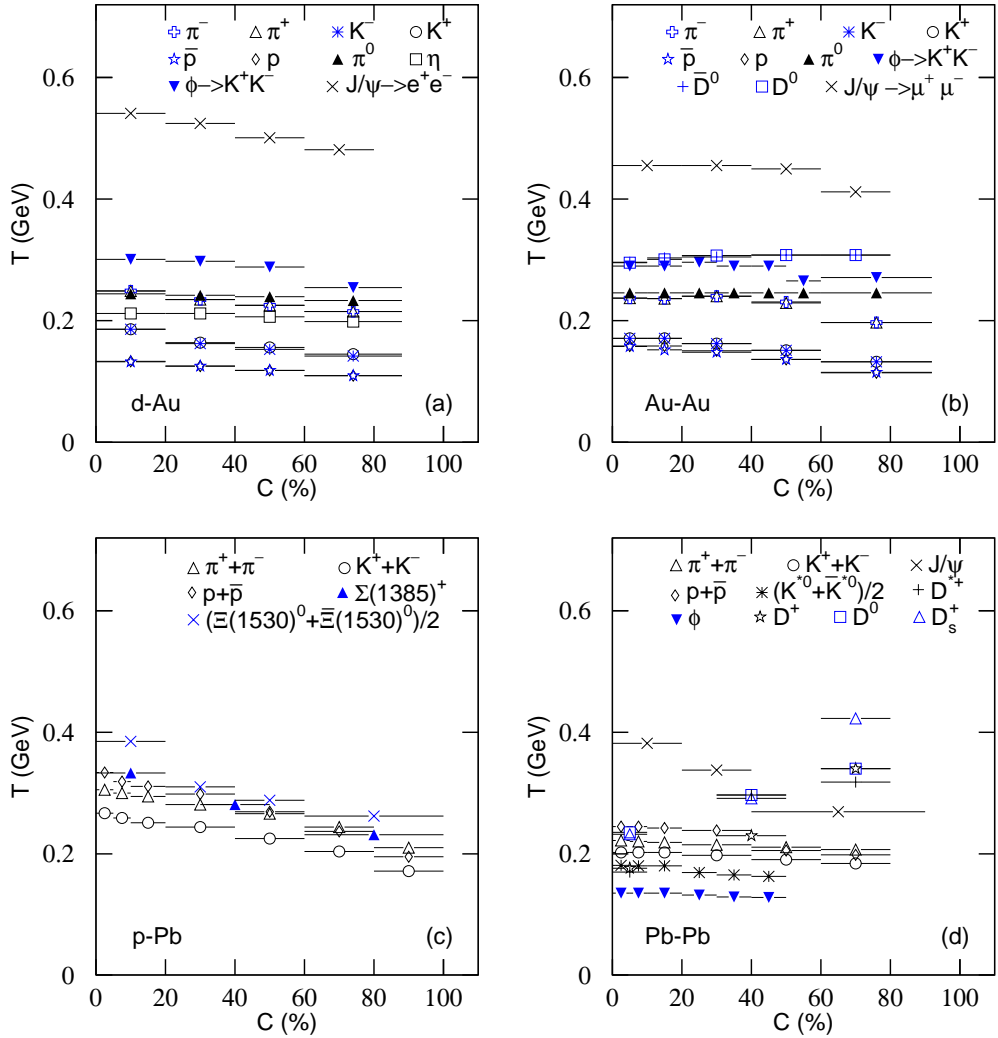}
\end{center}
\caption{\small Dependence of effective temperature $T$ on
centrality $C$ in (a) $d$-Au, (b) Au-Au, (c) $p$-Pb, and (d) Pb-Pb
collisions at the RHIC or LHC. The obtained values of parameter
$T$ corresponding to identified particles are extracted from the
experimental transverse momentum spectra and listed in Tables
2--5.}
\end{figure*}

\begin{figure*}[!htb]
\begin{center}
\includegraphics[width=13.0cm]{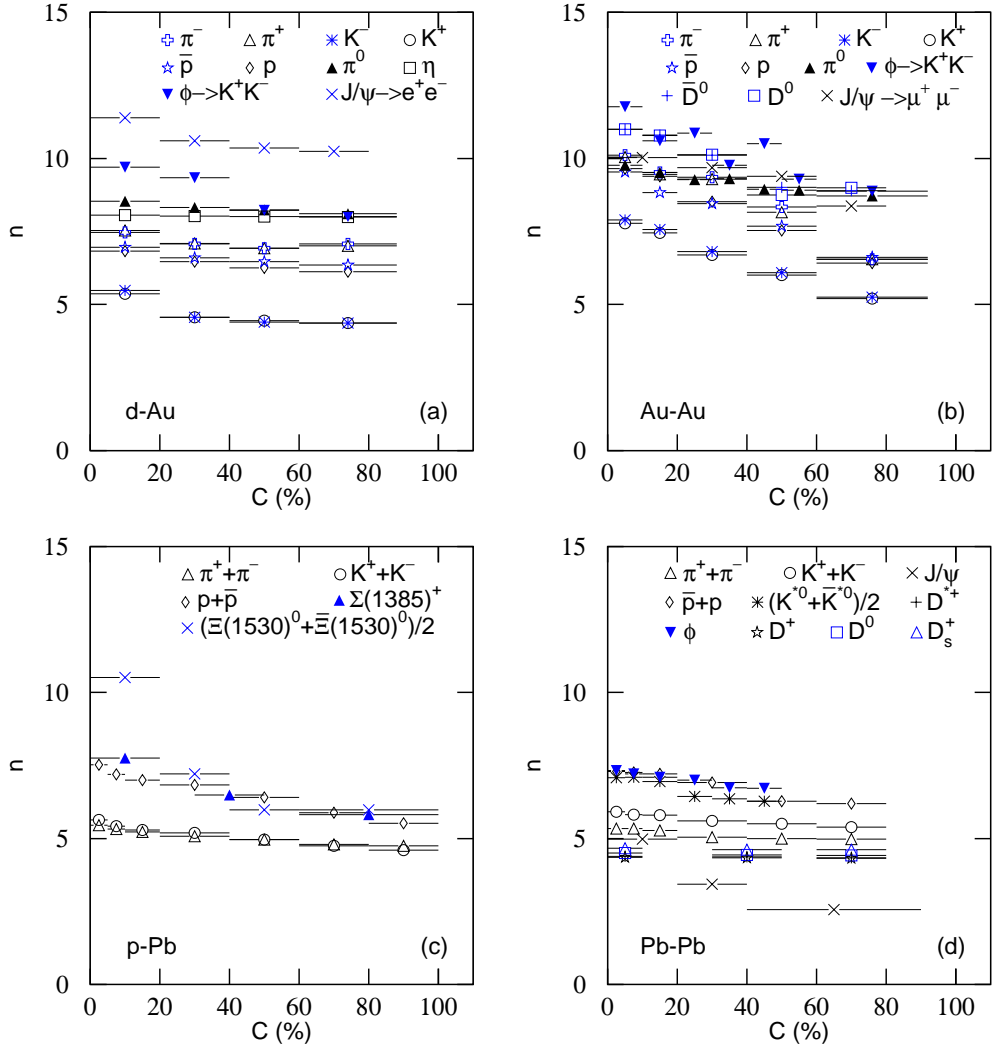}
\end{center}
\caption{\small Dependence of parameter $n$ on centrality $C$ in
(a) $d$-Au, (b) Au-Au, (c) $p$-Pb, and (d) Pb-Pb collisions at the
RHIC or LHC. The obtained parameter values corresponding to
identified particles are cited from Tables 2--5.}
\end{figure*}

\begin{figure*}
\begin{center}
\includegraphics[width=13.0cm]{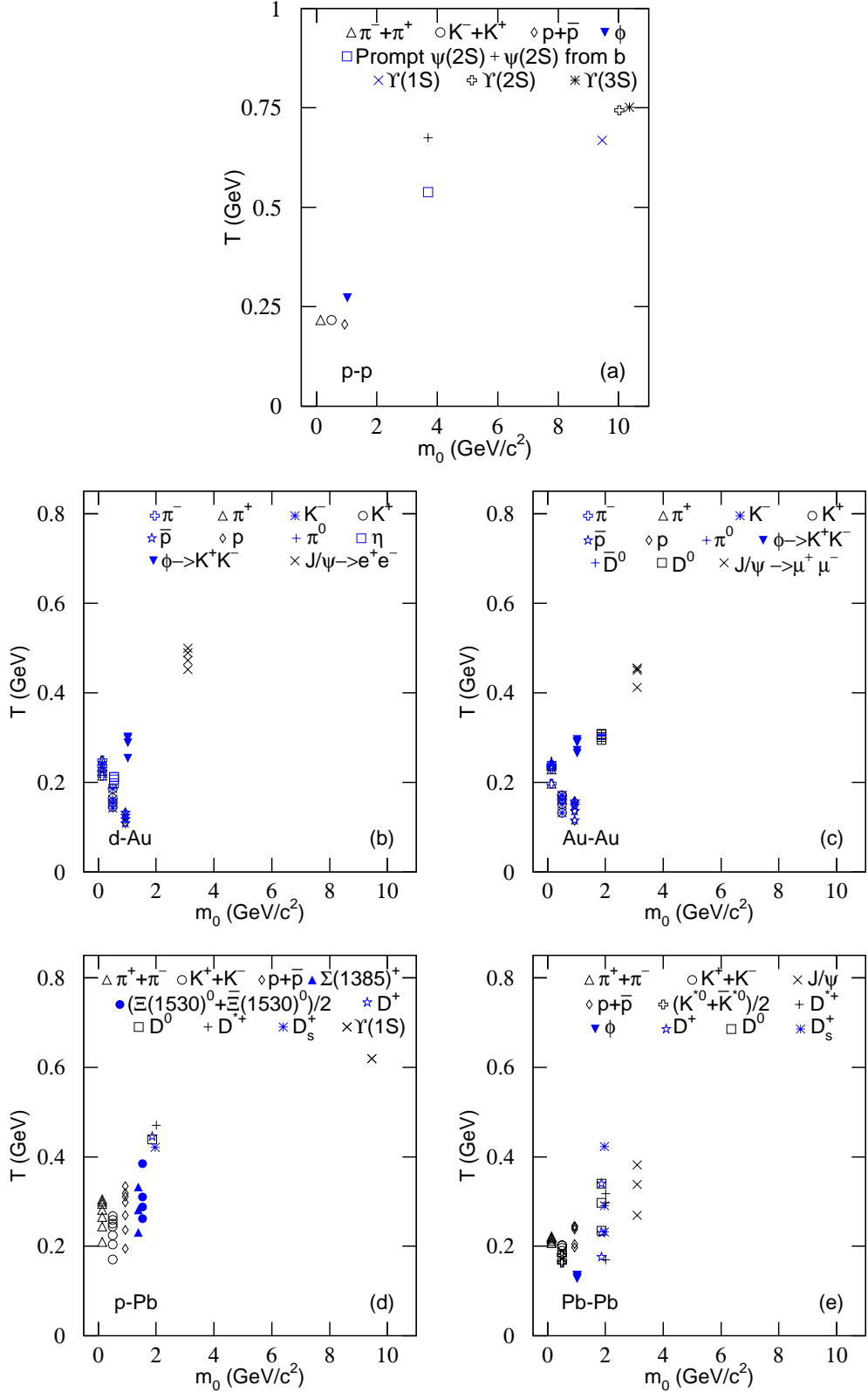}
\end{center}\vspace{-0.5cm}
\caption{\small Dependence of effective temperature $T$ on rest
mass $m_{0}$ of hadrons produced in (a) $p$-$p$, (b) $d$-Au, (c)
Au-Au, (d) $p$-Pb, and (e) Pb-Pb collisions at the RHIC or LHC.
The obtained values of parameter $T$ corresponding to identified
particles are extracted from the experimental transverse momentum
spectra and listed in Tables 1--5. The free parameter is extracted
at the quark level, though the dependence of free parameter on
rest mass of hadron is given.}
\end{figure*}

\begin{figure*}
\begin{center}
\includegraphics[width=13.0cm]{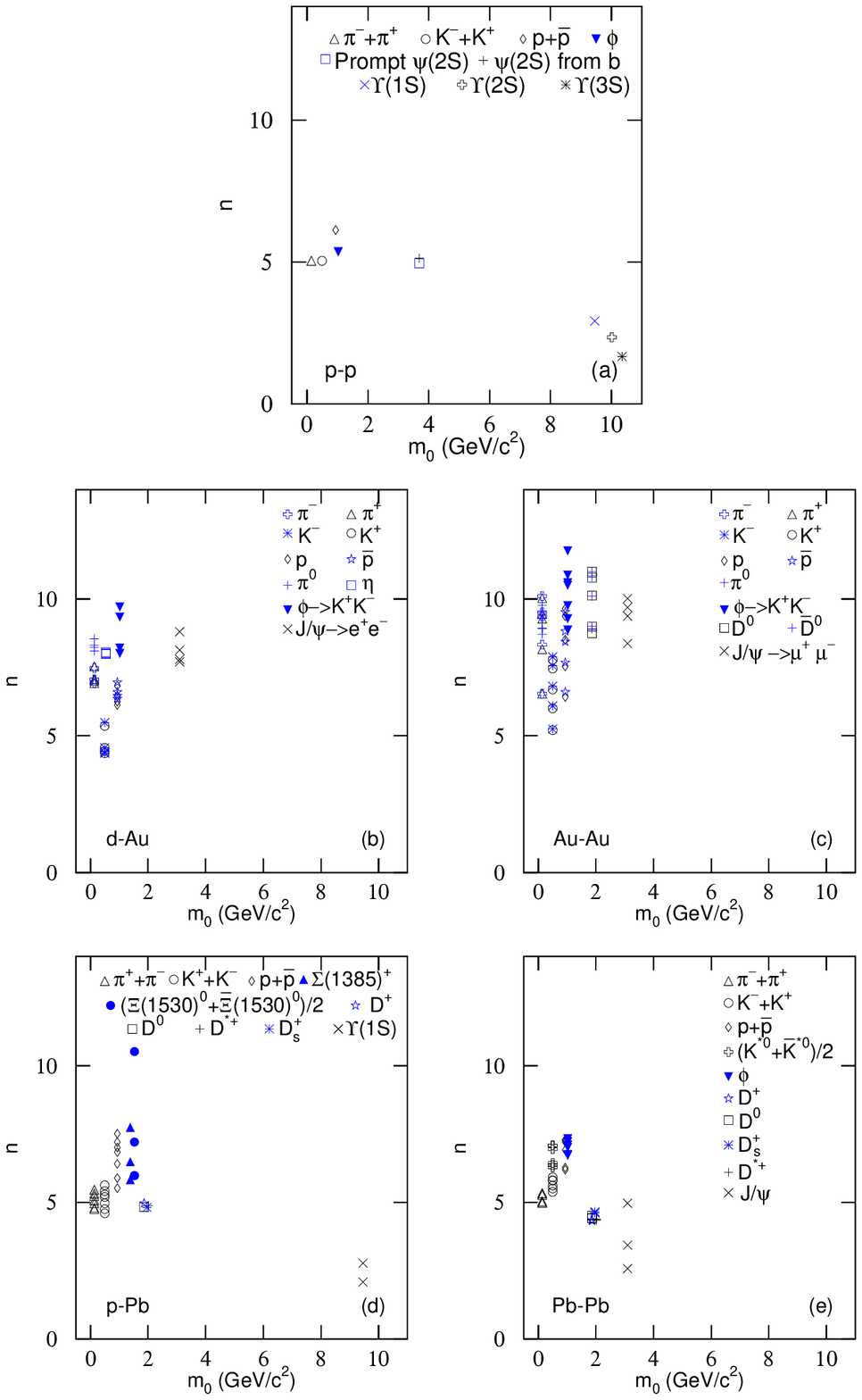}
\end{center}\vspace{-0.5cm}
\caption{\small Dependence of parameter $n$ on rest mass $m_{0}$
of hadrons in (a) $p$-$p$, (b) $d$-Au, (c) Au-Au, (d) $p$-Pb, and
(e) Pb-Pb collisions at the RHIC or LHC. The obtained parameter
values corresponding to identified particles are cited from Tables
1--5. The free parameter is extracted at the quark level, though
the dependence of free parameter on rest mass of hadron is given.}
\end{figure*}

\begin{figure*}
\begin{center}
\includegraphics[width=13.0cm]{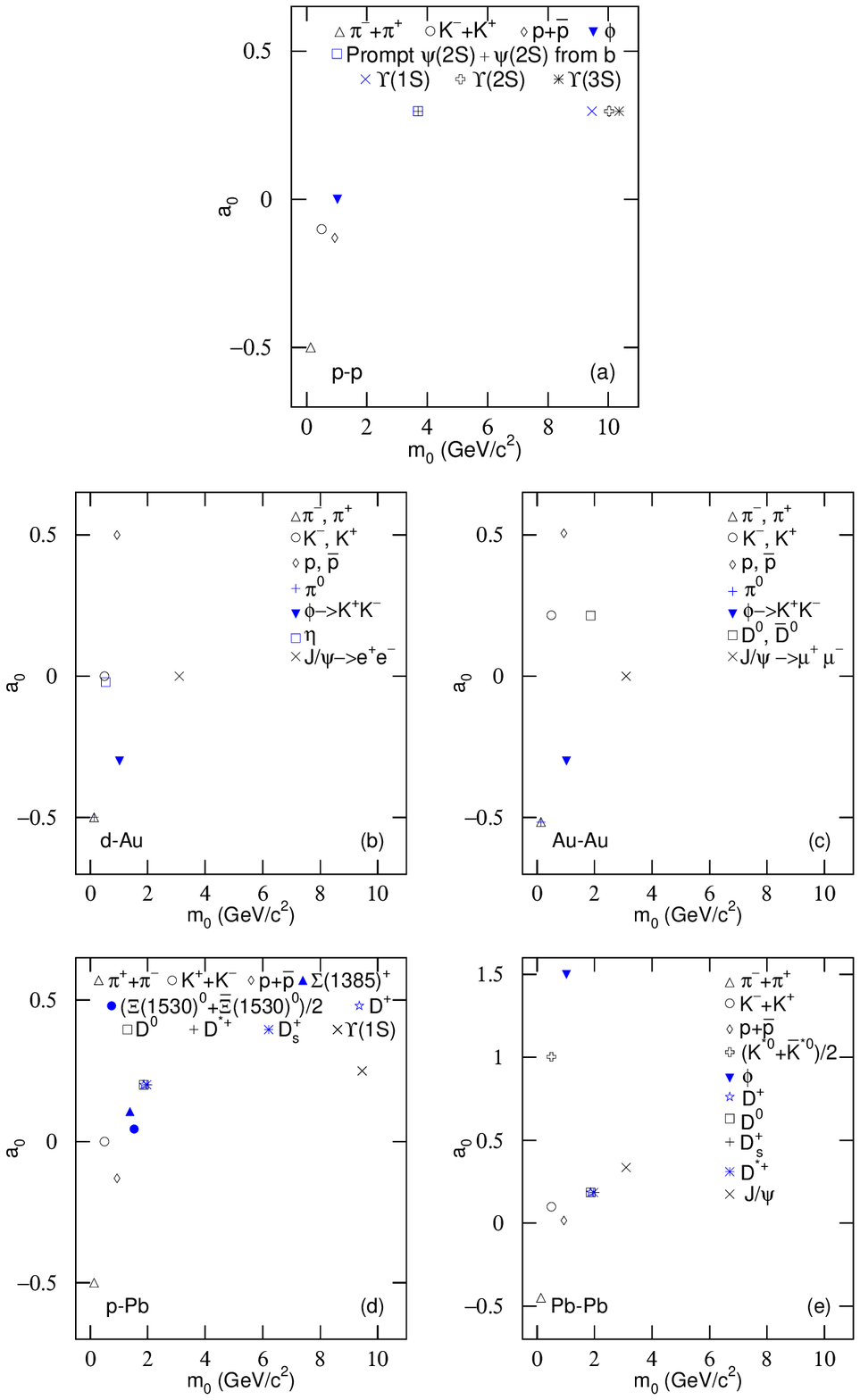}
\end{center}\vspace{-0.5cm}
\caption{\small Dependence of parameter $a_{0}$ on rest mass
$m_{0}$ of hadrons in (a) $p$-$p$, (b) $d$-Au, (c) Au-Au, (d)
$p$-Pb, and (e) Pb-Pb collisions at the RHIC or LHC. The obtained
parameter values corresponding to identified particles are cited
from Tables 1--5. The free parameter is extracted at the quark
level, though the dependence of free parameter on rest mass of
hadron is given.}
\end{figure*}

\begin{figure*}
\begin{center}
\includegraphics[width=13.0cm]{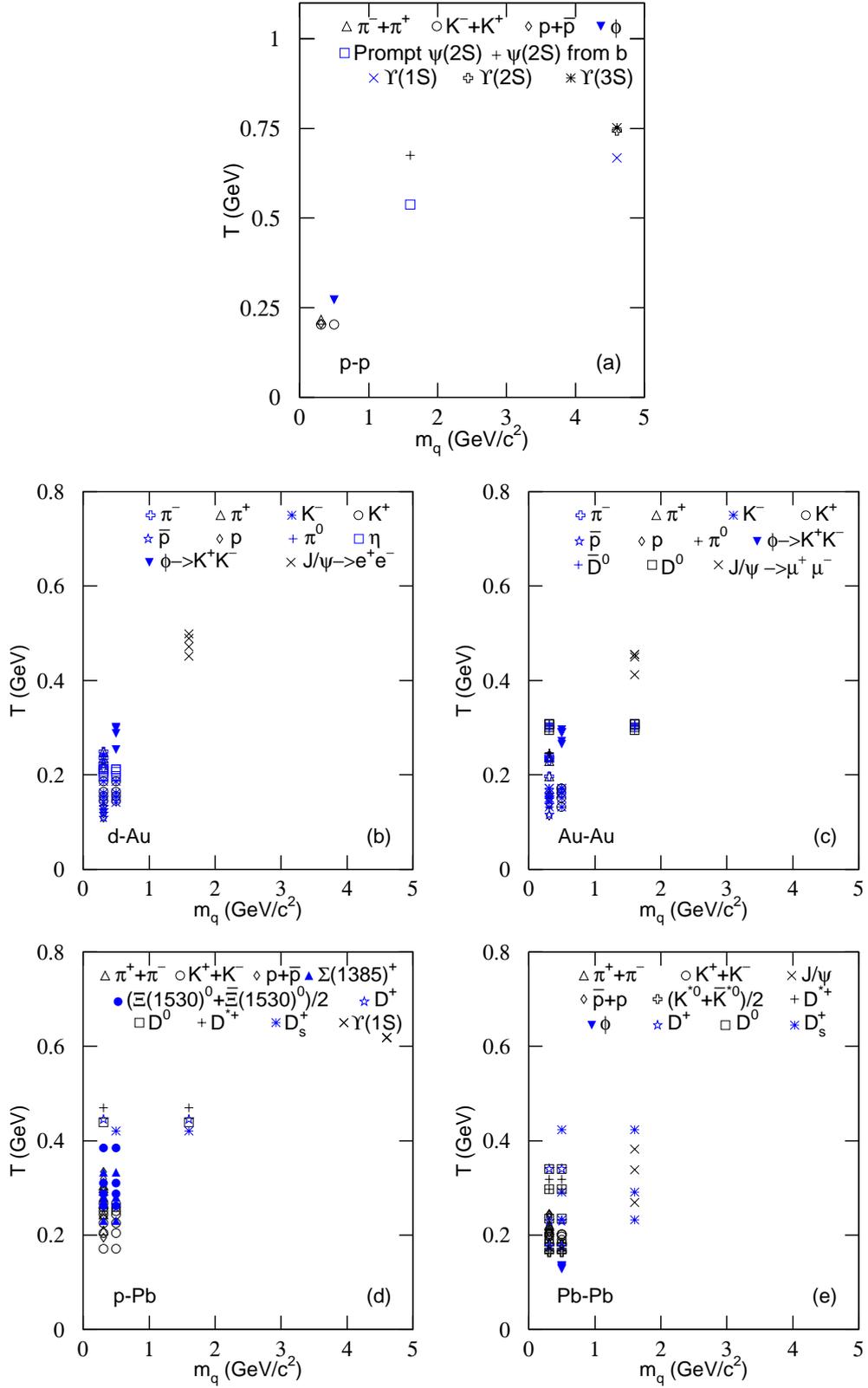}
\end{center}\vspace{-0.5cm}
\caption{\small Dependence of effective temperature $T$ on
constituent mass $m_{q}$ of quarks that make up hadrons produced
in (a) $p$-$p$, (b) $d$-Au, (c) Au-Au, (d) $p$-Pb, and (e) Pb-Pb
collisions at the RHIC or LHC. The obtained values of parameter
$a_0$ corresponding to various quarks are extracted from the
experimental transverse momentum spectra and listed in Tables
1--5.}
\end{figure*}

\begin{figure*}
\begin{center}
\includegraphics[width=13.0cm]{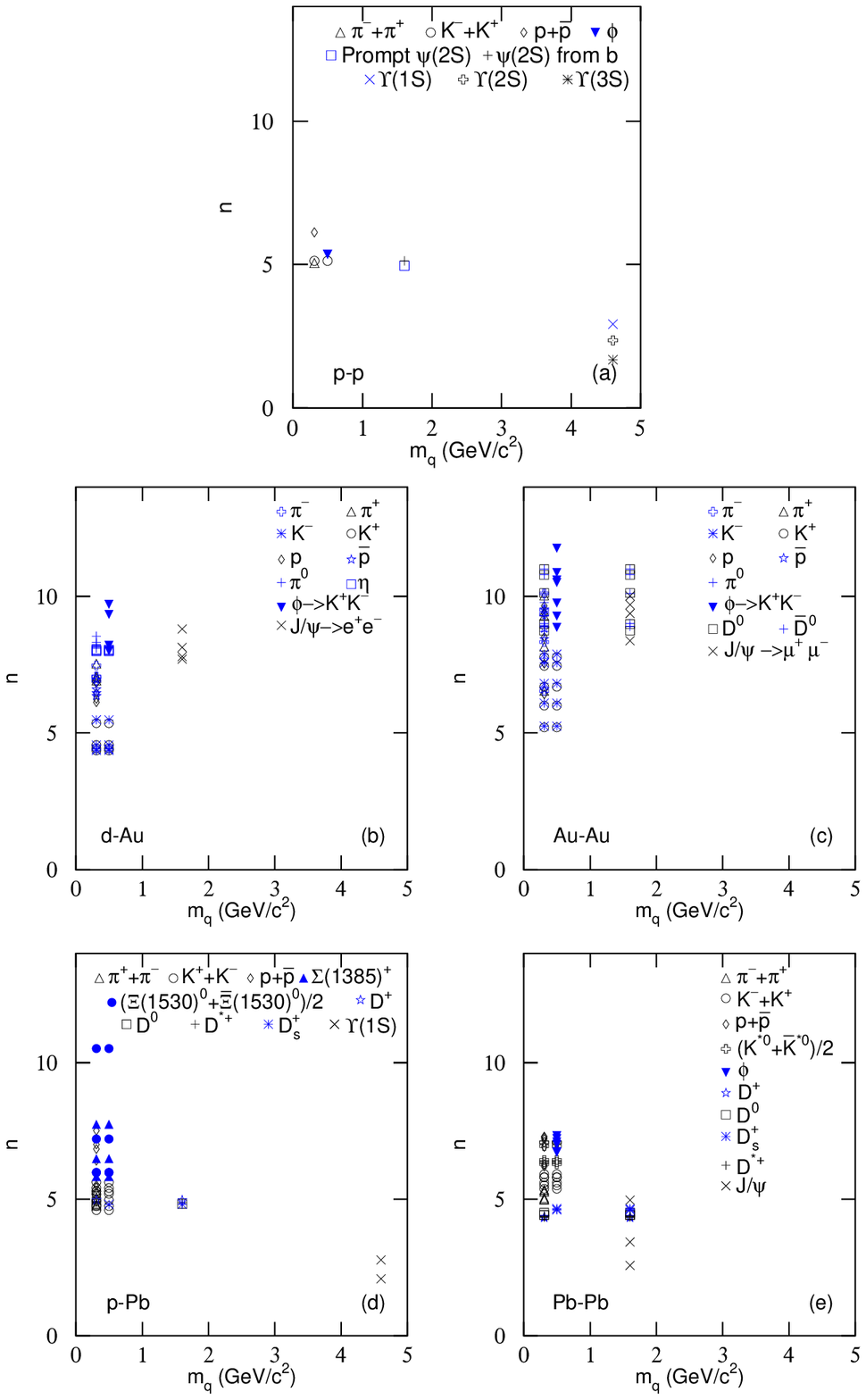}
\end{center}\vspace{-0.5cm}
\caption{\small Dependence of parameter $n$ on constituent mass
$m_{q}$ of quarks that make up hadrons produced in (a) $p$-$p$,
(b) $d$-Au, (c) Au-Au, (d) $p$-Pb, and (e) Pb-Pb collisions at the
RHIC or LHC. The obtained parameter values corresponding to
various quarks are cited from Tables 1--5.}
\end{figure*}

\begin{figure*}
\begin{center}
\includegraphics[width=13.0cm]{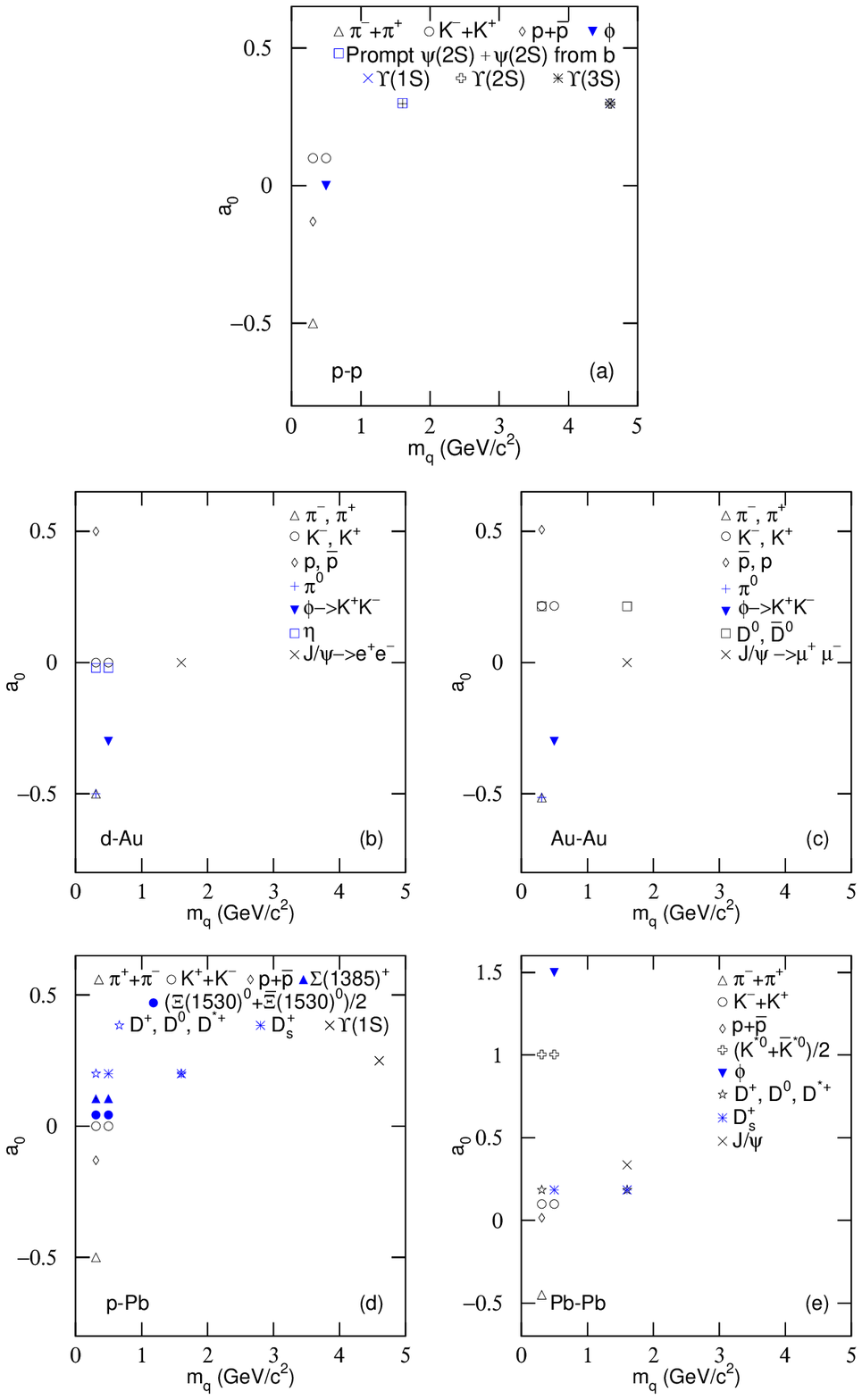}
\end{center}\vspace{-0.5cm}
\caption{\small Dependence of parameter $a_{0}$ on constituent
mass $m_{q}$ of quarks that make up hadrons produced in (a)
$p$-$p$, (b) $d$-Au, (c) Au-Au, (d) $p$-Pb, and (e) Pb-Pb
collisions at the RHIC or LHC. The obtained parameter values
corresponding to various quarks are cited from Tables 1--5.}
\end{figure*}

{\subsection{Dependence of parameters}}

We now analyze the dependences of effective temperature $T$,
entropy index-related $n$, and revised index $a_{0}$ on centrality
class $C$, rest mass $m_{0}$ of produced hadrons, and constituent
mass $m_{q}$ of participant quarks. The parameter values presented
in the following discussions are cited from Tables 1--5 which are
obtained from the fits to the experimental transverse momentum
spectra displayed in Figs. 1--5.

Figure 6 shows the centrality dependence of parameter $T$
extracted from the experimental data measured in (a) $d$-Au and
(b) Au-Au collisions at the RHIC, as well as (c) $p$-Pb and (d)
Pb-Pb collisions at the LHC, by international collaborations. One
can see that the effective temperature of the emission source
increases slowly with the increase of centrality from peripheral
to central collisions in most cases. This renders that more
collision energies are deposited in central collisions.

Figure 7 shows the dependence of parameter $n$ on centrality $C$,
where $n$ is extracted from the experimental data under different
conditions: (a) $d$-Au and (b) Au-Au collisions at the RHIC, as
well as (c) $p$-Pb and (d) Pb-Pb collisions at the LHC. In the
Tsallis statistics~\cite{13c,13d,13e,13f,13ff,33}, $n=1/(q-1)$,
where $q$ is the entropy index, refers to the equilibrium degree.
Generally, $q=1$ corresponds to an equilibrium state, and $q>1$
corresponds to a non-equilibrium. One can see that $n$ is very
large and increases slowly with the increase of centrality from
peripheral to central collisions, which means that $q$ is close to
1 and further closes to 1 slowly with the increase of centrality.
This study shows that the central collisions are closer to
equilibrium state than the peripheral collisions. This situation
is caused by more multiple scatterings in central collisions.

Now that we are talking about multiple scattering, we would like
to point out that the multiple scattering at quark level in view
of TP-like function used here differs from those at nucleonic
level. At quark level, less multiple scatterings and elastic
collisions should be happened due to very strong interactions
among quark, anti-quark, and gluon. At nucleonic level, more
multiple scatterings and elastic collisions should be happened due
to non-violent interactions between produced particle and nucleon.
Both the multiple scatterings at quark and nucleonic levels lead
to larger $p_{\rm T}$ and then wider spectra and larger $T$.
Although the influence of multiple scattering at quark level is
less than that at nucleonic level, the final effect is the sum of
two types of multiple scatterings.

From Tables 2--5 one can see that, for given particle and
collisions, the parameter $a_0$ is independent of centrality $C$.
So we have not displayed the figure about the dependence of
parameter $a_{0}$ on centrality $C$ to avoid trivialness. As a
parameter that characterizes the bending degree of the function
curve in the low $p_{\rm T}$ region, the fact that $a_0$ is
independent of $C$ implies that the spectator does not affect the
production of particles with low $p_{\rm T}$. Meanwhile, the size
of participant region does not affect the production at low
$p_{\rm T}$, too. The production at low $p_{\rm T}$ is mainly
determined by the individual nucleon-nucleon collision or two or
three participant quarks.

Figures 8, 9, and 10 show the trends of parameters $T$, $n$, and
$a_{0}$ with the rest mass $m_{0}$ of corresponding hadron,
respectively. In the three figures, panels (a)--(e) correspond to
the results of $p$-$p$, $d$-Au, Au-Au, $p$-Pb, and Pb-Pb
collisions, respectively. The results for various centrality
classes in nuclear collisions are displayed together without
distinction. In Fig. 8, the effective temperature $T$ is
positively correlated with the rest mass of hadron. This means
that the massive hadrons are produced at a higher temperature. In
Figs. 9(b) and 9(c), due to the small range of rest mass of
hadron, the parameter $n$ has no obvious dependence on rest mass
of hadron. In Figs. 9(a), 9(d), and 9(e), one can see that the
entropy index-related $n$ decreases with the increase of rest mass
of hadron. One can observe from Fig. 10 that as the rest mass of
hadron increases, the revised index $a_{0}$ tends to increase.

It is not easy to observe the trend of parameters with the
constituent quark mass $m_{q}$ of corresponding hadron in Tables
1--5. In order to observe it more intuitively, Figures 11, 12, and
13 show the dependences of parameters $T$, $n$, and $a_{0}$ on the
constituent quark mass $m_{q}$ of corresponding hadron,
respectively. In the three figures, panels (a)--(e) correspond to
the results of $p$-$p$, $d$-Au, Au-Au, $p$-Pb, and Pb-Pb
collisions, respectively. The results for various centrality
classes in nuclear collisions are displayed together without
distinction. One can clearly see that the correlation trend of
parameters with constituent quark mass of hadron is consistent
with that of parameters with rest mass of hadron. The parameters
$T$ and $a_{0}$ are positively correlated with the constituent
quark mass $m_{q}$. In Figs. 12(b) and 12(c), there is no obvious
change trend being observed due to the small constituent quark
mass range. In Figs. 12(a), 12(d), and 12(e), the parameter $n$
has a negative correlation with the constituent quark mass. It
should be pointed out that we have used the same symbol for the
parameter points of different quarks in the same hadron.
\\

{\subsection{More discussions}}

The trends of parameters $T$, $n$, and $a_{0}$ with the rest mass
$m_{0}$ of hadrons and with the constituent mass $m_{q}$ of quarks
are the same correspondingly. This phenomenon can be explained as
follows: the heavy hadrons are composed of heavy flavor quarks
with larger constituent mass, while the light hadrons correspond
to light flavor quarks with smaller constituent mass. The values
of effective temperature $T$ increase with the increase of mass
(rest mass of hadron or constituent mass of quarks), which can be
interpreted as the higher temperature is required for the
production of massive particles. The parameter $n$ decreases with
the increase of mass. This means that the heavier hadron
corresponds to the further away state from the equilibrium, though
the system is still close to the equilibrium.

Although there have existed a big number of publications which
went deeper in the transverse momentum
spectra~\cite{13d,34,34a,34b,35,35a,35b,36,36a,36b,37,37a,37b,38,38a,38b,38c},
especially for the analysis on the dependence of related
parameters on centrality and mass, we have applied another unique
Tsallis model to study the mentioned spectra. In our treatment,
for the model itself, different participant quark for given
particle has the same form of probability density function with
the same value of parameter. However, it was proposed to
contribute to the transverse momentum differently due to different
mass. This mass dependent transverse momentum spectrum is a
natural result.

Comparing with the fits of Tsallis-inspired models (not exactly
the same model) in other
publications~\cite{13d,34,34a,34b,35,35a,35b,36,36a,36b,37,37a,37b,38,38a,38b,38c},
the present work studies the spectra at the quark level and uses a
more flexible treatment for the spectra in very low transverse
momentum region. The related spectra studied in other publications
can also be fitted by the present model. In fact, not only for the
spectra with exponential form in low transverse momentum region
but also for the spectra with inverse power-law form (from
perturbative QCD) in medium and high transverse momentum region,
the present model can fit different spectra flexibly~\cite{13}.
Even for the spectra of jets, the present model can be used
approximately~\cite{50}.

There is connection between Eq. (1) and the well known inverse
power-law or the Hagedorn function which is empirically inspired
by perturbative QCD~\cite{50b,50b2}. Let $a_0=1$ and at high
$p_{\rm T}$ (i.e. neglect $m_0$) in Eq. (1), we have $f(p_{\rm
T})=C_0p_{\rm T}(1+p_{\rm T}/nT)^{-n}$, which is the inverse
power-law or the Hagedorn function. Here $T$, $n$, and $C_0$ are
different from those in Eq. (1) due to the refit and
normalization. If we use $n=1/(q-1)$, a different $q$ from the
above discussion can be obtained. This simplification results in
the meanings of parameters in Eq. (1) to be more obvious.
Meanwhile, Eq. (1) also has a more solid foundation. To fit the
spectra better, the inverse power-law or Hagedorn function has
been revised by at least three
forms~\cite{20,50b3,50b4,50b5,50b7,50b8,50b9}, which is beyond the
focus of the present work and will not be discussed anymore.

In the model, although the effective temperature \(T\) in the
function is possibly not the physical temperature exactly, its
physical significance is obvious. In fact, \(T\) reflects the
average kinetic energy of considered quarks. Generally, \(T\)
contains the contributions of thermal motion and collective
flow~\cite{50a}. Only the former contributes to the physical
temperature. In the process of participant quarks taking part in
the microscopic collisions, the collective flow has not yet formed
because the macroscopic squeeze and expansion of the collision
system is delayed to show the flow effect. So, as a quantity
extracted at the quark level, \(T\) is mainly contributed by the
thermal motion. In other words, we may regard \(T\) as the
physical temperature approximately.

The value of $T$ extracted at the quark level is obviously larger
than the critical temperature (\(\approx 0.16\) GeV) of chemical
freeze-out. There are two reasons causing this difference. On the
one hand, $T$ contains the contribution of flow effect which
increases $T$~\cite{50a}. On the other hand, $T$ extracted at the
quark level describes the earlier collision system which stays at
higher excitation state with higher temperature than the stage of
chemical freeze-out. Besides, different models use different
temperature parameters which can be fitted (``measured") by
different functions (``thermometers"). So, the difference between
\(T\) and critical temperature is natural. Of course, a unified
``thermometer" is needed in high energy collisions. This issue is
beyond the focus of the present work and we will not discuss it
anymore.

It is possible to quantify the thermal and collective part of the
hadron spectra in terms of the parameters extracted. In fact, in
our previous works~\cite{50a1,50a2,50a3,50a4,50a5,50a6}, the
intercept in the relation of effective temperature versus rest
mass of hadron is regarded as the physics temperature of emission
source at the stage of kinetic freeze-out. The slope in the
relation of average transverse momentum versus average energy of
hadron in the source rest frame is regarded as the transverse flow
velocity. According to the obtained parameters, we may calculate
the average transverse momentum and average energy, and obtain the
physics temperature and transverse flow velocity. Based on the
mentioned method of intercept and slope, we have quantified the
thermal and collective part in our previous
works~\cite{50a1,50a2,50a3,50a4,50a5,50a6} at the hadronic level.

Compared to the Tsallis and Hagedorn descriptions, the present
work presents a more accurate description of the transverse
momentum spectra in wider range and for more particles. At the
quark level, we should consider to use the quantities related to
quark to quantify the thermal and collective part. Based on our
experience, at least three types of quarks, $u(d)$, $s$, and $c$,
are needed. More quarks may possibly result in the mass-dependent
differential kinetic freeze-out scenario~\cite{50a3}. At the top
RHIC or LHC energy, for symmetric INEL $p$-$p$ and central $A$-$A$
collisions, the physics temperature estimated by us increases from
0.12 to 0.19 GeV with the increase of system size due to more
energy deposition in large system. The transverse flow velocity
decreases from 0.10 to 0.05 $c$ with the increase of system size
due to large stopping of medium for quarks in large system.

The Tsallis non-extensive entropy parameter \(q\) or the entropy
index-related \(n\) is related to equilibrium degree. In most
cases, when the collision system emits light and heavy hadrons,
\(q\) is close to 1 or \(n\) is large. This means that the
collision system at the RHIC and LHC stays in an approximate
equilibrium state. In fact, when the collision system emits
leptons~\cite{13}, it also stays in an approximate equilibrium
state. However, when the collision system emits jets~\cite{50}, it
stays in a non-equilibrium state in some cases. We may say that
the concept of temperature is approximately applicable in most
cases in high energy collisions.

In central collisions, we have obtained a larger $T$ and $n$ than
those in peripheral collisions. This means that central collisions
stay in higher excitation state and are closer to equilibrium
state. This is caused by more energy deposition due to more
participant nucleons in central collisions. For heavy hadrons and
heavy flavor quarks, a larger $T$ and smaller $n$ are obtained due
to early emission. In fact, in the early emission process, the
collision system has higher temperature and has no enough time to
research to a closer equilibrium state. The early emission of
heavy hadrons (and heavy flavor quarks) is a natural result of
hydrodynamic behavior~\cite{51}. This is because that massive
particles are left in the process of system evolution.

The revised index \(a_0\) describes the winding degree of the
spectra in very low transverse momentum region. Generally,
\(a_0<0\) is a reflection of increment in yield of light particles
via resonance decay, while \(a_0>0\) is a reflection of decrement
in yield of heavy particles due to decay or absorption in hot and
dense medium in participant region. In very low transverse
momentum region, the present work shows that both the effects of
the generation of light hadrons via resonance decay and the decay
or absorption of heavy hadrons in hot and dense medium are
obvious. The present work also shows that $a_0$ is independent of
centrality. This reflects that the resonance generation and decay
or absorption happen in the participant region, but not in the
spectator region. We may neglect the cold (spectator) nuclear
effect in the study of transverse momentum spectra.

Apart from in medium effects which provide the environment of
resonance and absorption, heavy ion collisions especially Pb-Pb
collisions also have additional final state effects such as
collective flow, recombination, and jet quenching in different
$p_{\rm T}$ regions which finally contributes to the hadron
spectra. These effects affect together the spectra in final state
by increasing or decreasing $p_{\rm T}$. In experiments, the
contributions of these effects cannot be absolutely distinguished.
We have uniformly described the different contributions by the
convolution of TP-like functions. Apart from $a_0$ which
determines the bending degree of curve in low $p_{\rm T}$ region
and then affects the slope of curve in medium and high $p_{\rm T}$
region, larger $T$ or smaller $n$ results in wider spectra, which
decreases the probability in low $p_{\rm T}$ region and increases
the probability in high $p_{\rm T}$ region.

Before summary and conclusions, we would like to point out that
Eqs. (3) or (5) can fit meanwhile the spectra in low and
medium-high transverse momentum regions which are contributed by
the soft excitation and hard scattering processes respectively.
For the spectra in low (or medium-high) transverse momentum region
only, Eqs. (3) or (5) is expected to fit naturally. In low energy
region such as in the RHIC Beam Energy Scan (BES) program or the
Super Proton Synchrotron (SPS) and its BES program, the spectra
are mainly contributed by the soft excitation process. The
contribution fraction of the hard scattering process at low energy
is less than that at the RHIC and LHC~\cite{13b}. In particular,
the contribution of hard process at low energy can be neglected in
most cases.

Eqs. (3) or (5) can fit the spectra at the RHIC BES and the SPS
authentically. In order to highlight the key points discussed
above, we show the comparisons of Eqs. (3) or (5) with the
experimental data measured at the SPS and RHIC BES in the Appendix
A. As examples, only INEL $p$-$p$, central Au-Au, and central
Pb-Pb collisions are included in Figs. 14--16, respectively.
\\

\clearpage

\end{multicols}
\begin{table*}
{\small Table 2. Values of $T$, $n$, $a_0$, $N_0$, $\chi^2$, and
ndof corresponding to the curves in Fig. 2. All the free
parameters are extracted at the quark level. \vspace{-0.5cm}
\begin{center}
\begin{tabular} {cccccccccccc}\\ \hline\hline
Particle & Centrality & $T$ (GeV) & $n$ & $a_{0}$ & $N_0$ & $\chi^2$/ndof \\
(quark structure) & (\%) &  &  &  &  & \\
\hline
                        & $0-20$     & $0.249\pm0.001$ & $7.466\pm0.026$ & $-0.500\pm0.001$ & $6.784\pm0.029$ & 47/20\\
$\pi^-$                 & $20-40$    & $0.235\pm0.001$ & $7.066\pm0.026$ & $-0.500\pm0.002$ & $5.218\pm0.028$ & 34/20\\
($d\bar u$)             & $40-60$    & $0.225\pm0.001$ & $6.924\pm0.023$ & $-0.500\pm0.002$ & $3.674\pm0.011$ & 54/20\\
                        & $60-88$    & $0.215\pm0.001$ & $7.064\pm0.025$ & $-0.500\pm0.002$ & $1.838\pm0.011$ & 81/20\\
                        & $0-100$    & $0.235\pm0.001$ & $7.154\pm0.022$ & $-0.500\pm0.001$ & $3.749\pm0.016$ & 64/20\\
\hline
                        & $0-20$     & $0.249\pm0.001$ & $7.537\pm0.022$ & $-0.500\pm0.002$ & $6.881\pm0.023$ & 56/20\\
$\pi^+$                 & $20-40$    & $0.235\pm0.001$ & $7.080\pm0.023$ & $-0.500\pm0.001$ & $5.276\pm0.028$ & 42/20\\
($u\bar d$)             & $40-60$    & $0.225\pm0.001$ & $6.922\pm0.023$ & $-0.500\pm0.002$ & $3.720\pm0.017$ & 50/20\\
                        & $60-88$    & $0.215\pm0.001$ & $7.007\pm0.022$ & $-0.500\pm0.002$ & $1.850\pm0.011$ & 99/20\\
                        & $0-100$    & $0.235\pm0.001$ & $7.159\pm0.015$ & $-0.500\pm0.001$ & $3.790\pm0.011$ & 53/20\\
\hline
                        & $0-20$     & $0.186\pm0.001$ & $5.482\pm0.018$ & $0.000\pm0.002$ & $0.906\pm0.007$ & 80/17\\
$K^-$                   & $20-40$    & $0.162\pm0.001$ & $4.560\pm0.021$ & $0.000\pm0.005$ & $0.677\pm0.003$ & 85/17\\
($s\bar u$)             & $40-60$    & $0.153\pm0.001$ & $4.403\pm0.020$ & $0.000\pm0.004$ & $0.459\pm0.003$ & 70/17\\
                        & $60-88$    & $0.142\pm0.001$ & $4.365\pm0.022$ & $0.000\pm0.003$ & $0.225\pm0.002$ & 47/17\\
                        & $0-100$    & $0.169\pm0.001$ & $4.873\pm0.021$ & $0.000\pm0.003$ & $0.475\pm0.002$ & 129/17\\
\hline
                        & $0-20$     & $0.186\pm0.001$ & $5.360\pm0.024$ & $0.000\pm0.002$ & $0.948\pm0.013$ & 89/17\\
$K^+$                   & $20-40$    & $0.164\pm0.001$ & $4.562\pm0.021$ & $0.000\pm0.002$ & $0.706\pm0.002$ & 90/17\\
($u\bar s$)             & $40-60$    & $0.156\pm0.001$ & $4.445\pm0.022$ & $0.000\pm0.003$ & $0.481\pm0.003$ & 39/17\\
                        & $60-88$    & $0.145\pm0.001$ & $4.360\pm0.023$ & $0.000\pm0.003$ & $0.233\pm0.003$ & 43/17\\
                        & $0-100$    & $0.172\pm0.001$ & $4.887\pm0.019$ & $0.000\pm0.002$ & $0.497\pm0.002$ & 110/17\\
\hline
                        & $0-20$     & $0.132\pm0.001$ & $6.955\pm0.024$ & $0.500\pm0.001$ & $0.013\pm0.002$ & 67/20\\
$\bar p$                & $20-40$    & $0.125\pm0.001$ & $6.599\pm0.023$ & $0.500\pm0.002$ & $0.007\pm0.000$ & 47/20\\
($\bar u\bar u\bar d$)  & $40-60$    & $0.118\pm0.001$ & $6.461\pm0.023$ & $0.500\pm0.002$ & $0.003\pm0.001$ & 34/20\\
                        & $60-88$    & $0.109\pm0.001$ & $6.350\pm0.024$ & $0.500\pm0.003$ & $0.001\pm0.000$ & 42/20\\
                        & $0-100$    & $0.124\pm0.001$ & $6.622\pm0.015$ & $0.500\pm0.002$ & $0.003\pm0.001$ & 65/20\\
\hline
                        & $0-20$     & $0.133\pm0.001$ & $6.831\pm0.015$ & $0.500\pm0.002$ & $0.023\pm0.002$ & 144/20\\
$p$                     & $20-40$    & $0.125\pm0.001$ & $6.460\pm0.019$ & $0.500\pm0.003$ & $0.012\pm0.001$ & 100/20\\
($uud$)                 & $40-60$    & $0.118\pm0.001$ & $6.252\pm0.023$ & $0.500\pm0.003$ & $0.005\pm0.001$ & 54/20\\
                        & $60-88$    & $0.109\pm0.001$ & $6.119\pm0.023$ & $0.500\pm0.002$ & $0.001\pm0.001$ & 76/20\\
                        & $0-100$    & $0.124\pm0.001$ & $6.441\pm0.021$ & $0.500\pm0.003$ & $0.006\pm0.001$ & 164/20\\
\hline
                               & $0-20$      & $0.244\pm0.001$ & $8.539\pm0.022$ & $-0.500\pm0.003$ & $9.873\pm0.072$ & 4/17\\
$\pi^0$                        & $20-40$     & $0.242\pm0.001$ & $8.312\pm0.020$ & $-0.500\pm0.002$ & $4.442\pm0.036$ & 11/17\\
($(u\bar u-d\bar d)/\sqrt{2}$) & $40-60$     & $0.239\pm0.001$ & $8.236\pm0.022$ & $-0.500\pm0.003$ & $4.600\pm0.036$ & 10/17\\
                               & $60-88$     & $0.233\pm0.001$ & $8.109\pm0.029$ & $-0.500\pm0.001$ & $2.262\pm0.021$ & 6/17\\
\hline
                        & $0-20$      & $0.301\pm0.002$ & $9.704\pm0.035$ & $-0.300\pm0.005$ & $(1.372\pm0.046)\times10^{-1}$ & 6/6\\
$\phi$                  & $20-40$     & $0.298\pm0.002$ & $9.337\pm0.032$ & $-0.300\pm0.003$ & $(9.028\pm0.072)\times10^{-2}$ & 6/6\\
($s\bar s$)             & $40-60$     & $0.288\pm0.002$ & $8.215\pm0.034$ & $-0.300\pm0.005$ & $(4.816\pm0.108)\times10^{-2}$ & 5/6\\
                        & $60-88$     & $0.254\pm0.003$ & $8.012\pm0.037$ & $-0.300\pm0.003$ & $(3.248\pm0.056)\times10^{-2}$ & 1/6\\
\hline
                                            & $0-20$      & $0.212\pm0.001$ & $8.061\pm0.025$ & $-0.021\pm0.002$ & $0.527\pm0.003$ & 6/9\\
$\eta$: $\eta_{q}$, $\eta_{s}$              & $20-40$     & $0.212\pm0.001$ & $8.023\pm0.019$ & $-0.021\pm0.003$ & $0.377\pm0.003$ & 5/9\\
($(u\bar u$+$d\bar d)/\sqrt{2}$, $s\bar s$) & $40-60$     & $0.206\pm0.001$ & $8.001\pm0.020$ & $-0.021\pm0.003$ & $0.272\pm0.003$ & 3/9\\
                                            & $60-88$     & $0.198\pm0.001$ & $7.992\pm0.020$ & $-0.021\pm0.003$ & $0.154\pm0.001$ & 8/8\\
\hline
                                      & $0-20$      & $0.541\pm0.002$ & $11.391\pm0.023$ & $0.000\pm0.004$ & $(8.807\pm0.088)\times10^{-6}$ & 8/9\\
$J\psi \longrightarrow e^{+}e^{-}$    & $20-40$     & $0.525\pm0.002$ & $10.602\pm0.030$ & $0.000\pm0.002$ & $(6.606\pm0.052)\times10^{-6}$ & 10/9\\
($c\bar c$)                           & $40-60$     & $0.501\pm0.001$ & $10.352\pm0.028$ & $0.000\pm0.002$ & $(4.404\pm0.013)\times10^{-6}$ & 11/7\\
                                      & $60-88$     & $0.481\pm0.001$ & $10.244\pm0.022$ & $0.000\pm0.002$ & $(2.204\pm0.009)\times10^{-6}$ & 4/7\\
\hline
\end{tabular}%
\end{center}}
\end{table*}

\begin{table*}[!htb]
{\small Table 3. Values of $T$, $n$, $a_0$, $N_0$, $\chi^2$, and
ndof corresponding to the curves in Fig. 3. All the free
parameters are extracted at the quark level.} \vspace{-0.5cm}
{\scriptsize
\begin{center}
\begin{tabular} {cccccccccccc}\\ \hline\hline
Particle & Centrality & $T$ (GeV) & $n$ & $a_{0}$ & $N_0$ & $\chi^2$/ndof \\
(quark structure) & (\%) &  &  &  &  & \\
\hline
                        & $0-10$     & $0.237\pm0.001$ & $10.110\pm0.023$ & $-0.515\pm0.002$ & $300.364\pm1.722$ & 125/22\\
$\pi^-$                 & $10-20$    & $0.236\pm0.001$ & $9.518 \pm0.021$ & $-0.515\pm0.002$ & $205.734\pm1.149$ & 90/22\\
($d\bar u$)             & $20-40$    & $0.240\pm0.001$ & $9.359 \pm0.033$ & $-0.515\pm0.001$ & $109.609\pm0.865$ & 131/22\\
                        & $40-60$    & $0.231\pm0.001$ & $8.336 \pm0.024$ & $-0.515\pm0.002$ & $20.677 \pm0.402$ & 128/22\\
                        & $60-92$    & $0.197\pm0.001$ & $6.539 \pm0.022$ & $-0.515\pm0.001$ & $8.938  \pm0.053$ & 104/22\\
\hline
                        & $0-10$     & $0.237\pm0.001$ & $10.040\pm0.027$ & $-0.515\pm0.002$ & $298.691\pm1.723$ & 168/22\\
$\pi^+$                 & $10-20$    & $0.236\pm0.001$ & $9.457 \pm0.026$ & $-0.515\pm0.002$ & $204.044\pm0.977$ & 109/22\\
($u\bar d$)             & $20-40$    & $0.240\pm0.001$ & $9.288 \pm0.032$ & $-0.515\pm0.001$ & $108.477\pm0.577$ & 166/22\\
                        & $40-60$    & $0.229\pm0.001$ & $8.159 \pm0.021$ & $-0.515\pm0.001$ & $39.768 \pm0.386$ & 116/22\\
                        & $60-92$    & $0.197\pm0.001$ & $6.541 \pm0.024$ & $-0.515\pm0.002$ & $8.938  \pm0.064$ & 103/22\\
\hline
                        & $0-10$     & $0.171\pm0.001$ & $7.887\pm0.013$ & $0.215\pm0.002$ & $39.205\pm0.142$ & 91/18\\
$K^-$                   & $10-20$    & $0.171\pm0.001$ & $7.563\pm0.022$ & $0.215\pm0.001$ & $26.138\pm0.085$ & 65/18\\
($s\bar u$)             & $20-40$    & $0.162\pm0.001$ & $6.814\pm0.023$ & $0.215\pm0.002$ & $14.428\pm0.071$ & 141/18\\
                        & $40-60$    & $0.151\pm0.001$ & $6.092\pm0.018$ & $0.215\pm0.002$ & $5.184 \pm0.028$ & 123/18\\
                        & $60-92$    & $0.132\pm0.001$ & $5.249\pm0.021$ & $0.215\pm0.002$ & $0.973 \pm0.013$ & 220/18\\
\hline
                        & $0-10$     & $0.171\pm0.001$ & $7.785\pm0.022$ & $0.215\pm0.002$ & $41.274\pm0.214$ & 49/18\\
$K^+$                   & $10-20$    & $0.171\pm0.001$ & $7.453\pm0.023$ & $0.215\pm0.002$ & $27.651\pm0.129$ & 28/18\\
($u\bar s$)             & $20-40$    & $0.162\pm0.001$ & $6.689\pm0.021$ & $0.215\pm0.003$ & $15.162\pm0.021$ & 55/18\\
                        & $40-60$    & $0.151\pm0.001$ & $6.005\pm0.017$ & $0.215\pm0.002$ & $5.480 \pm0.021$ & 86/18\\
                        & $60-92$    & $0.132\pm0.001$ & $5.207\pm0.015$ & $0.215\pm0.001$ & $1.041 \pm0.007$ & 129/18\\
\hline
                        & $0-10$     & $0.157\pm0.001$ & $9.528\pm0.017$ & $0.505\pm0.003$ & $18.229\pm0.091$ & 181/22\\
$\bar p$                & $10-20$    & $0.152\pm0.001$ & $8.828\pm0.018$ & $0.505\pm0.004$ & $8.639 \pm0.066$ & 251/22\\
($\bar u\bar u\bar d$)  & $20-40$    & $0.148\pm0.001$ & $8.445\pm0.016$ & $0.505\pm0.003$ & $2.617 \pm0.024$ & 281/22\\
                        & $40-60$    & $0.136\pm0.001$ & $7.677\pm0.020$ & $0.505\pm0.002$ & $0.372 \pm0.012$ & 289/22\\
                        & $60-92$    & $0.115\pm0.001$ & $6.605\pm0.019$ & $0.505\pm0.002$ & $0.015 \pm0.002$ & 339/22\\
\hline
                        & $0-10$     & $0.158\pm0.001$ & $9.651\pm0.002$ & $0.505\pm0.003$ & $33.974\pm0.083$ & 278/22\\
$p$                     & $10-20$    & $0.158\pm0.001$ & $9.381\pm0.013$ & $0.505\pm0.002$ & $15.180\pm0.091$ & 157/22\\
($uud$)                 & $20-40$    & $0.150\pm0.000$ & $8.520\pm0.015$ & $0.505\pm0.004$ & $4.617 \pm0.015$ & 292/22\\
                        & $40-60$    & $0.136\pm0.001$ & $7.528\pm0.019$ & $0.505\pm0.003$ & $0.644 \pm0.008$ & 254/22\\
                        & $60-92$    & $0.114\pm0.001$ & $6.410\pm0.011$ & $0.505\pm0.004$ & $0.025 \pm0.004$ & 391/22\\
\hline
                               & $0-10$      & $0.246\pm0.001$ & $9.770\pm0.026$ & $-0.515\pm0.003$ & $254.832\pm3.642$ & 14/14\\
                               & $10-20$     & $0.246\pm0.001$ & $9.523\pm0.027$ & $-0.515\pm0.003$ & $195.664\pm5.278$ & 8/13\\
                               & $20-30$     & $0.246\pm0.001$ & $9.278\pm0.022$ & $-0.515\pm0.002$ & $134.019\pm2.262$ & 5/11\\
$\pi^0$                        & $30-40$     & $0.246\pm0.001$ & $9.299\pm0.020$ & $-0.515\pm0.002$ & $94.639 \pm4.524$ & 10/11\\
($(u\bar u-d\bar d)/\sqrt{2}$) & $40-50$     & $0.246\pm0.001$ & $8.940\pm0.022$ & $-0.515\pm0.003$ & $55.949 \pm0.377$ & 3/10\\
                               & $50-60$     & $0.246\pm0.001$ & $8.909\pm0.029$ & $-0.515\pm0.002$ & $30.701 \pm0.226$ & 2/10\\
                               & $60-92$     & $0.246\pm0.001$ & $8.710\pm0.030$ & $-0.515\pm0.002$ & $8.050  \pm0.150$ & 3/10\\
                               & $0-92$      & $0.246\pm0.001$ & $9.133\pm0.027$ & $-0.515\pm0.004$ & $83.333 \pm2.262$ & 19/18\\
\hline
                        & $0-10$      & $0.290\pm0.002$ & $11.765\pm0.034$ & $-0.300\pm0.003$ & $9.694\pm0.073$ & 7/6\\
                        & $10-20$     & $0.290\pm0.002$ & $10.593\pm0.035$ & $-0.300\pm0.003$ & $6.954\pm0.113$ & 6/6\\
                        & $20-30$     & $0.290\pm0.002$ & $10.869\pm0.040$ & $-0.300\pm0.004$ & $4.832\pm0.073$ & 6/6\\
$\phi$                  & $30-40$     & $0.290\pm0.003$ & $9.763 \pm0.031$ & $-0.300\pm0.003$ & $3.072\pm0.160$ & 11/6\\
($s\bar s$)             & $40-50$     & $0.290\pm0.004$ & $10.500\pm0.030$ & $-0.300\pm0.003$ & $1.996\pm0.079$ & 10/6\\
                        & $50-60$     & $0.265\pm0.002$ & $9.282 \pm0.026$ & $-0.300\pm0.005$ & $1.271\pm0.055$ & 10/6\\
                        & $60-92$     & $0.271\pm0.002$ & $8.874 \pm0.024$ & $-0.300\pm0.004$ & $0.257\pm0.010$ & 14/6\\
                        & $0-92$      & $0.291\pm0.002$ & $10.932\pm0.025$ & $-0.300\pm0.003$ & $3.041\pm0.113$ & 19/6\\
\hline
                        & $0-10 $     & $0.295\pm0.001$ & $10.991\pm0.026$ & $0.214\pm0.003$ & $1.621\pm0.012$ & 3/7\\
$D^0$                   & $10-20$     & $0.301\pm0.001$ & $10.786\pm0.028$ & $0.214\pm0.003$ & $1.005\pm0.008$ & 8/7\\
($c\bar u$)             & $20-40$     & $0.307\pm0.001$ & $10.129\pm0.021$ & $0.214\pm0.004$ & $0.502\pm0.003$ & 11/7\\
                        & $40-60$     & $0.308\pm0.001$ & $8.746 \pm0.022$ & $0.214\pm0.002$ & $0.150\pm0.001$ & 2/7\\
                        & $60-80$     & $0.308\pm0.001$ & $8.993 \pm0.025$ & $0.214\pm0.002$ & $0.036\pm0.001$ & 5/6\\
\hline
                        & $0-10 $     & $0.296\pm0.001$ & $10.999\pm0.022$ & $0.214\pm0.003$ & $1.696\pm0.013$ & 5/7\\
$\bar D^0$              & $10-20$     & $0.303\pm0.001$ & $10.801\pm0.021$ & $0.214\pm0.002$ & $1.081\pm0.010$ & 4/7\\
($u\bar c$)             & $20-40$     & $0.305\pm0.001$ & $10.104\pm0.025$ & $0.214\pm0.003$ & $0.553\pm0.005$ & 8/7\\
                        & $40-60$     & $0.308\pm0.001$ & $9.001 \pm0.020$ & $0.214\pm0.004$ & $0.166\pm0.001$ & 4/6\\
                        & $60-80$     & $0.308\pm0.001$ & $8.892 \pm0.021$ & $0.214\pm0.003$ & $0.034\pm0.001$ & 6/6\\
\hline
                        & $0-20 $     & $0.455\pm0.001$ & $10.020\pm0.001$ & $0.000\pm0.002$ & $(3.266\pm0.063)\times10^{-4}$ & 5/5\\
$J/ \psi$               & $20-40$     & $0.455\pm0.002$ & $9.689 \pm0.002$ & $0.000\pm0.002$ & $(1.633\pm0.031)\times10^{-4}$ & 1/4\\
($c\bar c$)             & $40-60$     & $0.450\pm0.002$ & $9.381 \pm0.002$ & $0.000\pm0.003$ & $(1.256\pm0.013)\times10^{-4}$ & 5/4\\
                        & $60-80$     & $0.412\pm0.003$ & $8.375 \pm0.003$ & $0.000\pm0.002$ & $(1.633\pm0.003)\times10^{-4}$ & 2/2\\
                        & $0-80$      & $0.449\pm0.001$ & $9.575 \pm0.001$ & $0.000\pm0.003$ & $(1.381\pm0.019)\times10^{-4}$ & 1/5\\
\hline
\end{tabular}%
\end{center}}
\end{table*}

\begin{table*}[!htb]
{\small Table 4. Values of $T$, $n$, $a_0$, $N_0$, $\chi^2$, and
ndof corresponding to the curves in Fig. 4. All the free
parameters are extracted at the quark level. \vspace{-0.75cm}
\setlength{\tabcolsep}{1mm}
\begin{center}
\begin{tabular} {cccccccccccc}\\ \hline\hline
Particle & Centrality & $T$ (GeV) & $n$ & $a_{0}$ & $N_0$ & $\chi^2$/ndof \\
(quark structure) & (\%) &  &  &  &  & \\
\hline
                        & $0-5$       & $0.305\pm0.002$ & $5.461\pm0.020$ & $-0.500\pm0.002$ & $21.932\pm0.223$ & 48/54\\
                        & $5-10$      & $0.300\pm0.002$ & $5.328\pm0.022$ & $-0.500\pm0.004$ & $20.741\pm0.372$ & 51/54\\
                        & $10-20$     & $0.294\pm0.003$ & $5.231\pm0.023$ & $-0.500\pm0.004$ & $14.987\pm0.191$ & 49/54\\
$\pi^++\pi^-$           & $20-40$     & $0.281\pm0.002$ & $5.079\pm0.031$ & $-0.500\pm0.005$ & $11.566\pm0.127$ & 56/54\\
($u\bar d$, $d\bar u$)  & $40-60$     & $0.266\pm0.002$ & $4.965\pm0.021$ & $-0.500\pm0.004$ & $8.129 \pm0.095$ & 59/54\\
                        & $60-80$     & $0.244\pm0.001$ & $4.807\pm0.023$ & $-0.500\pm0.001$ & $5.055 \pm0.064$ & 61/54\\
                        & $80-100$    & $0.210\pm0.002$ & $4.752\pm0.010$ & $-0.500\pm0.003$ & $2.239 \pm0.016$ & 67/54\\
                        & NSD         & $0.279\pm0.002$ & $5.092\pm0.022$ & $-0.500\pm0.004$ & $8.509 \pm0.079$ & 37/54\\
\hline
                        & $0-5$       & $0.267\pm0.002$ & $5.635\pm0.023$ & $0.000\pm0.003$ & $2.983 \pm0.223$ & 20/47\\
                        & $5-10$      & $0.259\pm0.001$ & $5.421\pm0.040$ & $0.000\pm0.004$ & $2.375 \pm0.372$ & 18/47\\
                        & $10-20$     & $0.251\pm0.001$ & $5.292\pm0.022$ & $0.000\pm0.005$ & $1.990 \pm0.013$ & 11/47\\
$K^++K^-$               & $20-40$     & $0.244\pm0.002$ & $5.201\pm0.019$ & $0.000\pm0.004$ & $1.511 \pm0.013$ & 12/47\\
($u\bar s$, $s\bar u$)  & $40-60$     & $0.225\pm0.001$ & $4.972\pm0.018$ & $0.000\pm0.003$ & $1.035 \pm0.013$ & 18/47\\
                        & $60-80$     & $0.204\pm0.002$ & $4.752\pm0.023$ & $0.000\pm0.002$ & $0.616 \pm0.003$ & 34/47\\
                        & $80-100$    & $0.171\pm0.002$ & $4.610\pm0.010$ & $0.000\pm0.003$ & $0.261 \pm0.003$ & 47/47\\
                        & NSD         & $0.238\pm0.001$ & $5.158\pm0.020$ & $0.000\pm0.003$ & $1.121 \pm0.013$ & 11/47\\
\hline
                              & $0-5$       & $0.334\pm0.002$ & $7.522\pm0.029$ & $-0.130\pm0.002$ & $1.801 \pm0.018$ & 102/45\\
                              & $5-10$      & $0.319\pm0.002$ & $7.202\pm0.024$ & $-0.130\pm0.002$ & $1.521 \pm0.018$ & 80/45\\
                              & $10-20$     & $0.311\pm0.001$ & $7.003\pm0.031$ & $-0.130\pm0.002$ & $1.325 \pm0.008$ & 67/45\\
$p+\bar p$                    & $20-40$     & $0.298\pm0.002$ & $6.839\pm0.021$ & $-0.130\pm0.002$ & $1.051 \pm0.008$ & 29/45\\
($uud$, $\bar u\bar u\bar d$) & $40-60$     & $0.269\pm0.002$ & $6.414\pm0.038$ & $-0.130\pm0.002$ & $0.814 \pm0.008$ & 32/45\\
                              & $60-80$     & $0.237\pm0.001$ & $5.890\pm0.031$ & $-0.130\pm0.004$ & $0.553 \pm0.006$ & 13/45\\
                              & $80-100$    & $0.195\pm0.001$ & $5.520\pm0.025$ & $-0.130\pm0.004$ & $0.269 \pm0.002$ & 16/45\\
                              & NSD         & $0.292\pm0.002$ & $6.790\pm0.021$ & $-0.130\pm0.003$ & $0.801 \pm0.007$ & 33/45\\
\hline
                        & $0-20 $     & $0.336\pm0.002$ & $5.513\pm0.037$ & $-0.100\pm0.003$ & $(0.048 \pm0.226)\times10^{-3}$ & 16/2\\
$\sum(1385)^+$          & $20-60$     & $0.299\pm0.003$ & $5.595\pm0.040$ & $-0.100\pm0.004$ & $(0.028 \pm0.570)\times10^{-3}$ & 9/2\\
($uus$)                 & $60-100$    & $0.281\pm0.002$ & $6.009\pm0.050$ & $-0.100\pm0.004$ & $(0.007 \pm0.115)\times10^{-3}$ & 9/2\\
                        & NSD         & $0.329\pm0.002$ & $6.254\pm0.032$ & $-0.100\pm0.002$ & $(0.024 \pm0.567)\times10^{-3}$ & 4/2\\
\hline
                                    & $0-20 $     & $0.419\pm0.003$ & $9.896\pm0.025$ & $-0.100\pm0.002$ & $(13.262\pm0.265)\times10^{-3}$ & 5/6\\
$(\Xi(1530)^0+\bar\Xi(1530)^0)/2$   & $20-40$     & $0.375\pm0.002$ & $8.553\pm0.033$ & $-0.100\pm0.015$ & $(8.516 \pm0.266)\times10^{-3}$ & 1/6\\
($uss$, $\bar u\bar s\bar s$)       & $40-60$     & $0.358\pm0.002$ & $7.195\pm0.031$ & $-0.100\pm0.013$ & $(4.793 \pm0.106)\times10^{-3}$ & 4/6\\
                                    & $60-100$    & $0.331\pm0.002$ & $7.637\pm0.032$ & $-0.100\pm0.001$ & $(1.710 \pm0.011)\times10^{-3}$ & 2/6\\
                                    & NSD         & $0.377\pm0.002$ & $8.099\pm0.041$ & $-0.100\pm0.002$ & $(5.905 \pm0.005)\times10^{-3}$ & 2/6\\
\hline
$D^+$ ($c\bar d$)                           & MB  & $0.445\pm0.002$ & $4.969\pm0.021$ & $0.200\pm0.003$ & $(1.911\pm0.002)\times10^{-2}$ & 4/14\\
$D^0$ ($c\bar u$)                           & MB  & $0.439\pm0.001$ & $4.840\pm0.022$ & $0.200\pm0.003$ & $(3.995\pm0.006)\times10^{-2}$ & 4/15\\
$D^{*+}$ ($c\bar d$)                        & MB  & $0.470\pm0.001$ & $4.878\pm0.021$ & $0.200\pm0.002$ & $(1.701\pm0.004)\times10^{-2}$ & 2/13\\
$D_{s}^+$ ($c\bar s$)                       & MB  & $0.421\pm0.002$ & $4.816\pm0.024$ & $0.200\pm0.004$ & $(1.161\pm0.003)\times10^{-2}$ & 2/1\\
$\Upsilon(1S)$ ($b\bar b$) ($-4.46<y<2.96$) & MB  & $0.619\pm0.004$ & $2.778\pm0.025$ & $0.250\pm0.003$ & $(2.719\pm0.075)\times10^{-7}$ & 3/1\\
$\Upsilon(1S)$ ($b\bar b$) ($2.03<y<3.53$)  & MB  & $0.619\pm0.002$ & $2.098\pm0.023$ & $0.250\pm0.002$ & $(4.522 \pm0.090)\times10^{-7}$ & 5/1\\
\hline
\end{tabular}%
\end{center}}
\end{table*}

\begin{table*}[!htb]
{\small Table 5. Values of $T$, $n$, $a_0$, $N_0$, $\chi^2$, and
ndof corresponding to the curves in Fig. 5, where ``--" in the
last column denotes ndof = 0 and the corresponding curve is for
the eye guiding only. All the free parameters are extracted at the
quark level. \vspace{-0.5cm}
\begin{center}
\begin{tabular} {cccccccccccc}\\ \hline\hline
Particle & Centrality & $T$ (GeV) & $n$ & $a_{0}$ & $N_0$ & $\chi^2$/ndof \\
(quark structure) & (\%) &  &  &  &  & \\
\hline
                        & $0-5$       & $0.222\pm0.001$ & $5.343\pm0.023$ & $-0.450\pm0.003$ & $(2.307\pm0.021)\times10^{3}$ & 328/59\\
                        & $5-10$      & $0.220\pm0.002$ & $5.343\pm0.022$ & $-0.450\pm0.002$ & $(2.013\pm0.012)\times10^{3}$ & 255/59\\
$\pi^++\pi^-$           & $10-20$     & $0.219\pm0.001$ & $5.285\pm0.021$ & $-0.450\pm0.003$ & $(1.527\pm0.013)\times10^{3}$ & 228/59\\
($u\bar d$, $d\bar u$)  & $20-40$     & $0.215\pm0.001$ & $5.055\pm0.023$ & $-0.450\pm0.003$ & $(8.979\pm0.085)\times10^{2}$ & 162/59\\
                        & $40-60$     & $0.211\pm0.001$ & $5.005\pm0.021$ & $-0.450\pm0.003$ & $(3.404\pm0.022)\times10^{2}$ & 69/59\\
                        & $60-80$     & $0.207\pm0.001$ & $4.985\pm0.020$ & $-0.450\pm0.004$ & $(9.571\pm0.075)\times10^{1}$ & 20/59\\
\hline
                        & $0-5$       & $0.202\pm0.001$ & $5.923\pm0.024$ & $0.100\pm0.004$ & $(3.451 \pm0.033)\times10^{2}$ & 434/54\\
                        & $5-10$      & $0.202\pm0.002$ & $5.822\pm0.030$ & $0.100\pm0.002$ & $(2.682 \pm0.022)\times10^{2}$ & 751/54\\
$K^++K^-$               & $10-20$     & $0.202\pm0.002$ & $5.802\pm0.020$ & $0.100\pm0.003$ & $(2.225 \pm0.022)\times10^{2}$ & 309/54\\
($u\bar s$, $s\bar u$)  & $20-40$     & $0.197\pm0.001$ & $5.608\pm0.021$ & $0.100\pm0.004$ & $(1.271 \pm0.011)\times10^{2}$ & 208/54\\
                        & $40-60$     & $0.190\pm0.001$ & $5.505\pm0.021$ & $0.100\pm0.004$ & $(4.964 \pm0.045)\times10^{1}$ & 55/54\\
                        & $60-80$     & $0.184\pm0.001$ & $5.385\pm0.024$ & $0.100\pm0.004$ & $(1.279 \pm0.011)\times10^{1}$ & 26/54\\
\hline
                              & $0-5$       & $0.245\pm0.002$ & $7.296\pm0.022$ & $0.015\pm0.003$ & $(1.083 \pm0.012)\times10^{2}$ & 631/45\\
                              & $5-10$      & $0.245\pm0.001$ & $7.266\pm0.021$ & $0.015\pm0.003$ & $(9.408 \pm0.015)\times10^{1}$ & 519/45\\
$p+\bar p$                    & $10-20$     & $0.242\pm0.002$ & $7.216\pm0.030$ & $0.015\pm0.003$ & $(7.186 \pm0.115)\times10^{1}$ & 485/45\\
($uud$, $\bar u\bar u\bar d$) & $20-40$     & $0.238\pm0.002$ & $6.916\pm0.023$ & $0.015\pm0.004$ & $(3.946 \pm0.035)\times10^{1}$ & 345/45\\
                              & $40-60$     & $0.205\pm0.002$ & $6.276\pm0.024$  & $0.015\pm0.002$ & $(1.734 \pm0.012)\times10^{1}$ & 234/45\\
                              & $60-80$     & $0.198\pm0.002$ & $6.200\pm0.026$  & $0.015\pm0.004$ & $4.955 \pm0.048$ & 48/45\\
\hline
                         & $0-5$       & $0.181\pm0.002$ & $7.088\pm0.022$ & $1.003\pm0.003$ & $19.493\pm0.190$ & 25/9\\
                         & $5-10$      & $0.180\pm0.001$ & $7.093\pm0.024$ & $1.003\pm0.004$ & $15.100\pm0.317$ & 22/9\\
$(K^{*0}+\bar K^{*0})/2$ & $10-20$     & $0.180\pm0.002$ & $6.952\pm0.023$ & $1.003\pm0.005$ & $12.930\pm0.654$ & 43/9\\
($d\bar s$, $s\bar d$)   & $20-30$     & $0.169\pm0.002$ & $6.441\pm0.024$ & $1.003\pm0.005$ & $10.073\pm0.095$ & 24/9\\
                         & $30-40$     & $0.165\pm0.002$ & $6.361\pm0.025$ & $1.003\pm0.004$ & $7.144 \pm0.069$ & 10/9\\
                         & $40-50$     & $0.163\pm0.001$ & $6.284\pm0.024$ & $1.003\pm0.003$ & $4.491 \pm0.019$ & 6/9\\
\hline
                         & $0-5$       & $0.135\pm0.002$ & $7.328\pm0.016$ & $1.500\pm0.004$ & $12.354\pm0.089$ & 37/13\\
                         & $5-10$      & $0.135\pm0.001$ & $7.222\pm0.022$ & $1.500\pm0.005$ & $10.560\pm0.127$ & 41/13\\
$\phi$                   & $10-20$     & $0.135\pm0.002$ & $7.096\pm0.024$ & $1.500\pm0.003$ & $8.255\pm0.065$ & 33/13\\
($s\bar s$)              & $20-30$     & $0.132\pm0.001$ & $7.002\pm0.022$ & $1.500\pm0.004$ & $6.401\pm0.084$ & 24/13\\
                         & $30-40$     & $0.129\pm0.001$ & $6.746\pm0.020$ & $1.500\pm0.003$ & $3.905\pm0.072$ & 14/13\\
                         & $40-50$     & $0.128\pm0.001$ & $6.717\pm0.021$ & $1.500\pm0.003$ & $2.497\pm0.052$ & 9/13\\
\hline
$D^+$                    & $0-10$      & $0.176\pm0.001$ & $4.352\pm0.003$ & $0.185\pm0.001$ & $12.217\pm0.010$ & 4/8\\
($c\bar d$)              & $30-50$     & $0.230\pm0.001$ & $4.342\pm0.006$ & $0.185\pm0.002$ & $1.528 \pm0.010$ & 6/7\\
                         & $60-80$     & $0.340\pm0.002$ & $4.329\pm0.005$ & $0.185\pm0.002$ & $0.079 \pm0.001$ & 3/6\\
\hline
$D^0$                    & $0-10$      & $0.235\pm0.001$ & $4.505\pm0.003$ & $0.185\pm0.002$ & $12.364\pm0.059$ & 15/9\\
($c\bar u$)              & $30-50$     & $0.297\pm0.001$ & $4.442\pm0.002$ & $0.185\pm0.001$ & $1.606 \pm0.007$ & 3/8\\
                         & $60-80$     & $0.340\pm0.001$ & $4.421\pm0.002$ & $0.185\pm0.002$ & $0.166 \pm0.001$ & 7/8\\
\hline
$D^{*+}$                 & $0-10$      & $0.170\pm0.001$ & $4.396\pm0.002$ & $0.185\pm0.003$ & $16.500\pm0.030$ & 2/7\\
($c\bar d$)              & $30-50$     & $0.297\pm0.002$ & $4.367\pm0.005$ & $0.185\pm0.003$ & $0.754 \pm0.003$ & 4/7\\
                         & $60-80$     & $0.318\pm0.001$ & $4.346\pm0.004$ & $0.185\pm0.002$ & $0.108 \pm0.001$ & 2/6\\
\hline
$D_{s}^{+}$              & $0-10$      & $0.232\pm0.001$ & $4.670\pm0.002$ & $0.185\pm0.001$ & $6.074 \pm0.030$ & 0.1/--\\
($c\bar s$)              & $30-50$     & $0.291\pm0.001$ & $4.613\pm0.005$ & $0.185\pm0.002$ & $0.690 \pm0.003$ & 0.4/--\\
                         & $60-80$     & $0.423\pm0.001$ & $4.613\pm0.004$ & $0.185\pm0.002$ & $0.043 \pm0.001$ & 0.5/1\\
\hline
$J/\psi$                 & $0-20$      & $0.382\pm0.001$ & $4.975\pm0.004$ & $0.335\pm0.001$ & $(19.036\pm0.179)\times10^{-2}$ & 0.5/--\\
($c\bar c$)              & $20-40$     & $0.338\pm0.001$ & $3.445\pm0.005$ & $0.335\pm0.002$ & $(5.692 \pm0.044)\times10^{-2}$ & 1/--\\
                         & $40-90$     & $0.269\pm0.001$ & $2.575\pm0.005$ & $0.335\pm0.002$ & $(0.761 \pm0.058)\times10^{-2}$ & 9/--\\
\hline
\end{tabular}%
\end{center}}
\end{table*}
\begin{multicols}{2}

{\section{Summary and conclusions}}

We summarize here our main observations and conclusions.

(a) We have studied the transverse momentum spectra of different
hadrons produced in $p$-$p$, $d$-Au, Au-Au, $p$-Pb, and Pb-Pb
collisions at different center-of-mass energies. The experimental
data measured by the ALICE, CMS, LHCb, NA49, NA61/SHINE, PHENIX,
and STAR Collaborations are collected and analyzed. In the
framework of multisource thermal model at the quark level or the
participant quark model, the fitting results are approximately in
agreement with the experimental data.

(b) The TP-like function, i.e. the revised Tsallis--Pareto-type
function, is used to describe the amount or portion contributed by
each participant quark to the transverse momentum distribution of
hadrons. The transverse momentum distribution of mesons is the
convolution of two TP-like functions because the meson is composed
of two quarks. The transverse momentum distribution of baryons is
a convolution of three TP-like functions because the baryon is
composed of three quarks. Not only for meson spectra but also for
baryon spectra, the number of parameters is always four that
includes three free parameters and a normalization constant. All
free parameters are extracted at the quark level.

(c) For $d$-Au, Au-Au, $p$-Pb, and Pb-Pb collisions at high
energies, in most cases, the effective temperature $T$ and entropy
index-related $n$ extracted from the fits of transverse momentum
spectra of hadrons increase with increasing the centrality from
peripheral to central collisions. The revised index $a_0$ does not
change with centrality. Peripheral collisions are further away
from the equilibrium state, though the system in peripheral
collisions is still close to the equilibrium state.

(d) The correlation trends of parameters $T$, $n$, and $a_{0}$
with the rest mass $m_{0}$ of hadron are in agreement with those
of the mentioned parameters with the constituent mass $m_{q}$ of
quark, respectively. $T$ has a positive correlation trend with
mass. $n$ has no obvious trend or has a negative correlation with
mass. $a_{0}$ is positively related to mass. The production of
hadrons with larger $m_0$ requires a higher $T$, and the
corresponding state is further away from the equilibrium state,
though the collision system is still close to the equilibrium
state.

(e) High energy $p$-$p$ and nuclear ($p(d)$-$A$ or $A$-$A$)
collisions show similar behaviors in many aspects, indicating that
we may use the same function to fit uniformly the spectra from
both the collisions. These similarities exist commonly in high
energy collisions from small system ($p$-$p$ or $p(d)$-$A$) to
large one ($A$-$A$), implying that the similar participant or
contributor quarks take part in the production of given particles,
in which the remaining spectator quarks do not affect mainly the
particle spectrum.
\\

\noindent {\bf Acknowledgments} This work was supported by the
National Natural Science Foundation of China under Grant Nos.
12047571, 11575103, and 11947418, the China Scholarship Council
(Chinese Government Scholarship) under Grant No. 202008140170, the
Shanxi Provincial Innovative Foundation for Graduate Education
under Grant No. 2019SY053, the Scientific and Technological
Innovation Programs of Higher Education Institutions in Shanxi
(STIP) under Grant No. 201802017, the Shanxi Provincial Natural
Science Foundation under Grant No. 201901D111043, and the Fund for
Shanxi ``1331 Project" Key Subjects Construction.
\\
\\
{\bf Data availability statement} This manuscript has no
associated data or the data will not be deposited. [Authors'
comment: The data used to support the findings of this study are
included within the article and are cited at relevant places
within the text as references.]
\\
\\
{\bf Compliance with ethical standards}
\\
\\
{\bf Ethical approval} The authors declare that they are in
compliance with ethical standards regarding the content of this
paper.
\\
\\
{\bf Disclosure} The funding agencies have no role in the design
of the study; in the collection, analysis, or interpretation of
the data; in the writing of the manuscript, or in the decision to
publish the results.
\\
\\
{\bf Conflict of interest} The authors declare that there are no
conflicts of interest regarding the publication of this paper.
\\
\\

\noindent {\bf Appendix A. Comparison with data at low energy}

Figure 14 shows the transverse momentum spectra, $d^2N/dp_{\rm
T}dy$, of (a) $\pi^+$ and $\pi^-$, (b) $K^+$ and $K^-$, as well as
(c) $p$ and $\bar p$ with $0<y<0.2$ produced in INEL $p$-$p$
collisions at 6.3, 7.7, 8.8, 12.3, and 17.3 GeV. The symbols
represent the experimental data measured by the NA61/SHINE
Collaborations~\cite{70,70a} at the SPS. The curves are our fitted
results by using Eqs. (3) for mesons and (5) for baryons and the
related parameters are listed in Table 6. One can see that the
curves fit well the experimental data measured by the NA61/SHINE
Collaboration.

Figure 15 shows the transverse momentum spectra, $(1/2\pi p_{\rm
T})d^2N/dp_{\rm T}dy$, of (a) $\pi^+$ and $\pi^-$, (b) $K^+$ and
$K^-$, (c) $p$ and $\bar p$, (d) $K_S^0$, (e) $\Lambda$ and
$\bar\Lambda$, as well as (f) $\bar\Xi^+$ and $\Xi^-$ with
(a)--(c) $|y|<0.1$ and (d)--(f) $|y|<0.5$ produced in 0--5\% Au-Au
collisions at 7.7, 11.5, 19.6, 27, and 39 GeV. The symbols
represent the experimental data measured by the STAR
Collaborations~\cite{71,71a} at the RHIC BES. The curves are our
fitted results by using Eqs. (3) for mesons and (5) for baryons
and the related parameters are listed in Table 7. One can see that
the curves fit well the experimental data measured by the STAR
Collaboration.

\begin{figure*}[!htb]
\begin{center}
\includegraphics[width=13.0cm]{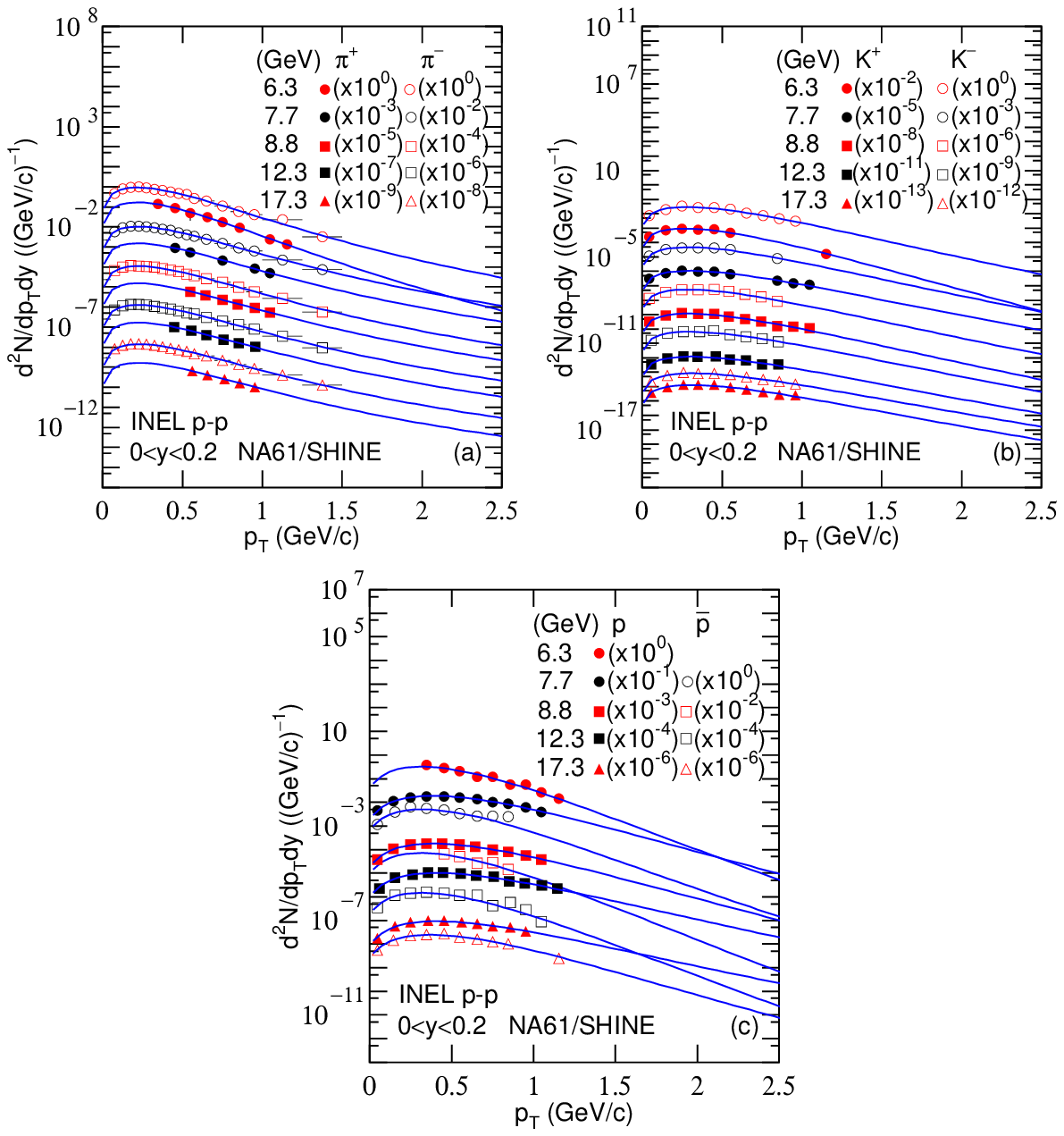}
\end{center}
\caption{\small The transverse momentum spectra of (a) $\pi^+$ and
$\pi^-$, (b) $K^+$ and $K^-$, as well as (c) $p$ and $\bar p$ with
$0<y<0.2$ produced in INEL $p$-$p$ collisions at 6.3, 7.7, 8.8,
12.3, and 17.3 GeV. The symbols represent the experimental data
measured by the NA61/SHINE Collaborations~\protect\cite{70,70a} at
the SPS and the curves are our fitted results by using Eqs. (3)
for mesons and (5) for baryons.}
\end{figure*}

\begin{figure*}
\begin{center}
\includegraphics[width=13.0cm]{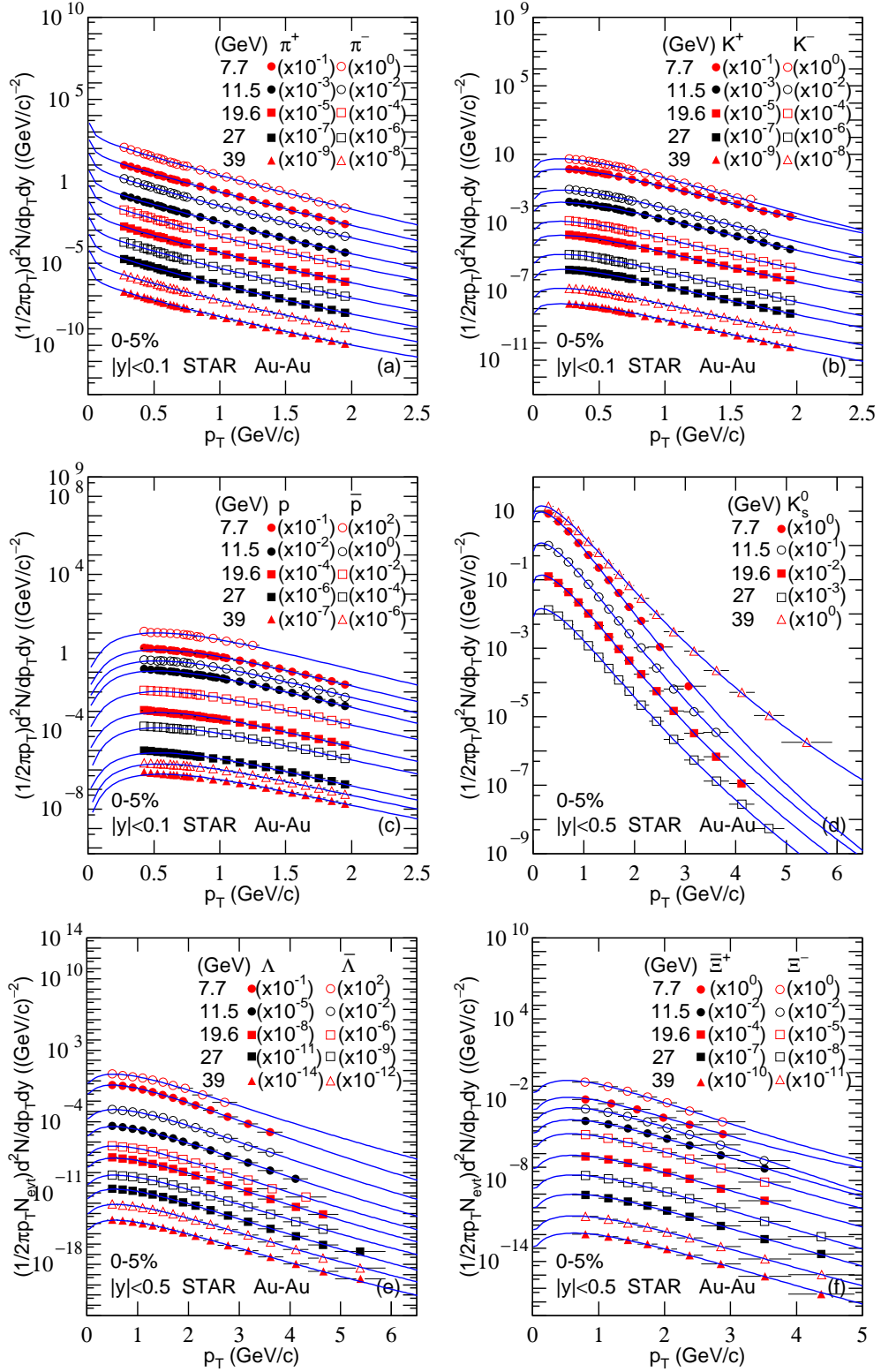}
\end{center}\vspace{-0.5cm}
\caption{\small The transverse momentum spectra of (a) $\pi^+$ and
$\pi^-$, (b) $K^+$ and $K^-$, (c) $p$ and $\bar p$, (d) $K_S^0$,
(e) $\Lambda$ and $\bar\Lambda$, as well as (f) $\bar\Xi^+$ and
$\Xi^-$ with (a)--(c) $|y|<0.1$ and (d)--(f) $|y|<0.5$ produced in
0--5\% Au-Au collisions at 7.7, 11.5, 19.6, 27, and 39 GeV. The
symbols represent the experimental data measured by the STAR
Collaborations~\protect\cite{71,71a} at the RHIC BES and the
curves are our fitted results by using Eqs. (3) for mesons and (5)
for baryons.}
\end{figure*}

\begin{figure*}
\begin{center}
\includegraphics[width=13.0cm]{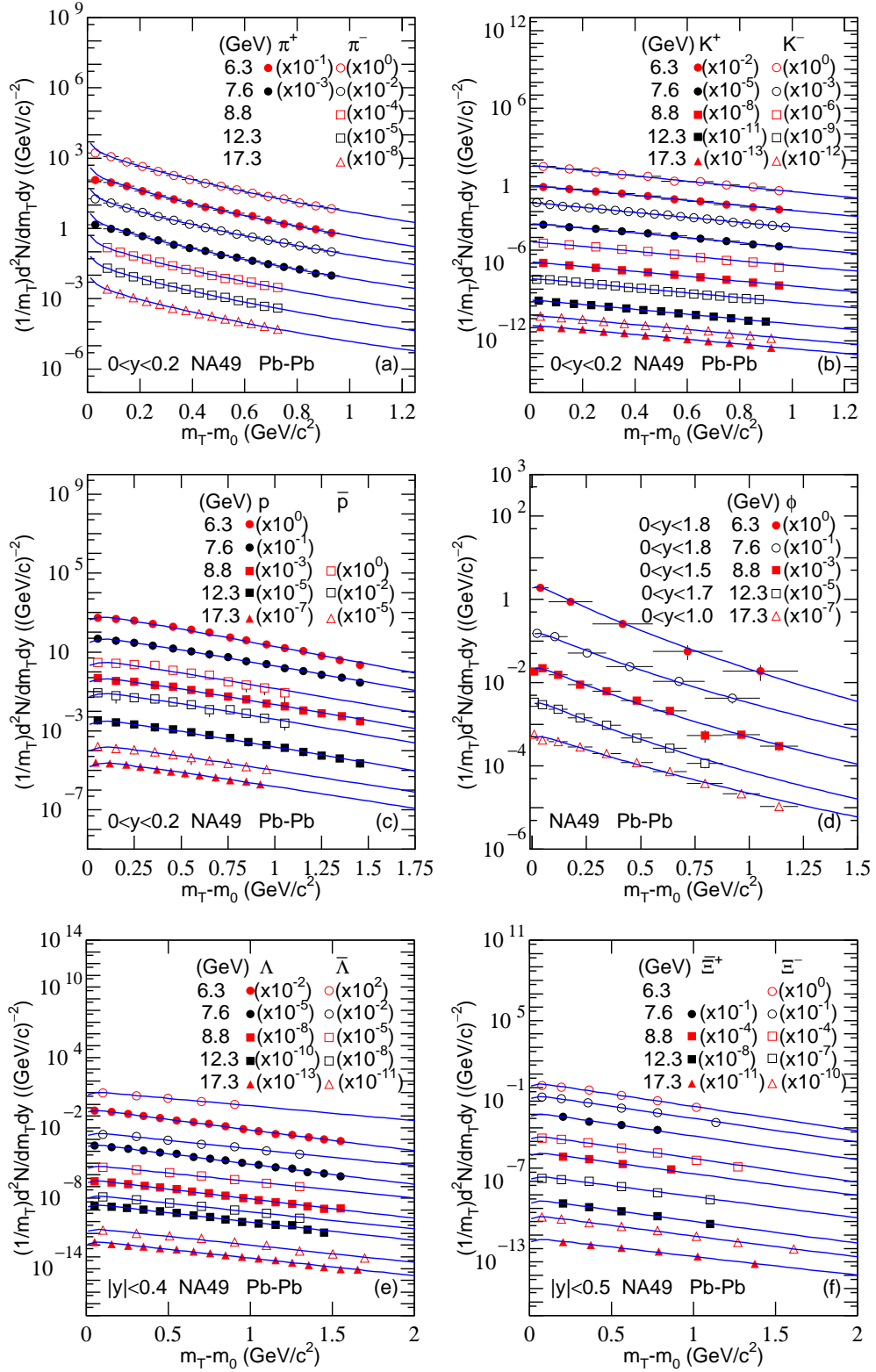}
\end{center}\vspace{-0.5cm}
\caption{\small The transverse mass spectra of (a) $\pi^+$ and
$\pi^-$, (b) $K^+$ and $K^-$, (c) $p$ and $\bar p$, (d) $\phi$,
(e) $\Lambda$ and $\bar\Lambda$, as well as (f) $\bar\Xi^+$ and
$\Xi^-$ with (a)--(c), (e), and (f) $0<y<0.2$, as well as (d)
$0<y<1.0$ to $0<y<1.8$ produced in central (0--5\% to 0--10\%)
Pb-Pb collisions at 6.3, 7.6, 8.8, 12.3, and 17.3 GeV. The symbols
represent the experimental data measured by the NA49
Collaborations~\protect\cite{72,72a,72b,72c,72d} at the SPS and
the curves are our fitted results by using Eqs. (3) for mesons and
(5) for baryons.}
\end{figure*}

\end{multicols}
\begin{table*}[!htb]
{\small Table 6. Values of $T$, $n$, $a_{0}$, $N_0$, $\chi^2$, and
ndof corresponding to the curves in Fig. 14. All the free
parameters are extracted at the quark level. \vspace{-0.5cm}
\begin{center}
\begin{tabular} {cccccccccccc}\\ \hline\hline
Particle & Energy & $T$ (GeV) & $n$ & $a_{0}$ & $N_0$ & $\chi^2$/ndof \\
(quark structure) & (GeV) &  &  &  &  & \\
\hline
                        & $6.3$       & $0.090\pm0.001$ & $9.965\pm0.040$ & $-0.03\pm0.003$ & $0.082\pm0.001$  & 7/14\\
$\pi^-$                 & $7.7$       & $0.092\pm0.001$ & $7.847\pm0.041$ & $-0.03\pm0.001$ & $0.095\pm0.001$  & 14/14\\
($d\bar u$)             & $8.8$       & $0.092\pm0.001$ & $7.857\pm0.043$ & $-0.03\pm0.002$ & $0.102\pm0.001$  & 24/14\\
                        & $12.3$      & $0.092\pm0.001$ & $7.212\pm0.048$ & $-0.03\pm0.002$ & $0.120\pm0.001$  & 23/14\\
                        & $17.3$      & $0.093\pm0.001$ & $6.978\pm0.052$ & $-0.03\pm0.001$ & $0.134\pm0.002$  & 13/14\\
\hline
                        & $6.3$       & $0.090\pm0.001$ & $15.667\pm0.037$ & $-0.03\pm0.002$ & $0.014\pm0.001$ & 0.1/3\\
$\pi^+$                 & $7.7$       & $0.091\pm0.001$ & $7.602\pm0.036$  & $-0.03\pm0.002$ & $0.142\pm0.001$ & 0.3/1\\
($u\bar d$)             & $8.8$       & $0.092\pm0.001$ & $7.444\pm0.035$  & $-0.03\pm0.003$ & $0.145\pm0.001$ & 0.1/2\\
                        & $12.3$      & $0.092\pm0.001$ & $6.949\pm0.051$  & $-0.03\pm0.002$ & $0.160\pm0.001$ & 0.2/2\\
                        & $17.3$      & $0.093\pm0.001$ & $6.908\pm0.034$  & $-0.03\pm0.001$ & $0.160\pm0.001$ & 0.1/1\\
\hline
                        & $6.3$     & $0.113\pm0.001$ & $23.368\pm0.053$ & $0.000\pm0.002$ & $(3.207\pm0.002)\times10^{-3}$ & 3/6\\
$K^-$                   & $7.7$     & $0.115\pm0.001$ & $21.249\pm0.045$ & $0.000\pm0.002$ & $(5.212\pm0.020)\times10^{-3}$ & 1/3\\
($s\bar u$)             & $8.8$     & $0.116\pm0.001$ & $21.921\pm0.043$ & $0.000\pm0.002$ & $(6.014\pm0.004)\times10^{-3}$ & 2/5\\
                        & $12.3$    & $0.125\pm0.002$ & $21.703\pm0.054$ & $0.000\pm0.002$ & $(7.417\pm0.023)\times10^{-3}$ & 2/5\\
                        & $17.3$    & $0.136\pm0.001$ & $18.997\pm0.052$ & $0.000\pm0.001$ & $(1.042\pm0.008)\times10^{-2}$ & 1/6\\
\hline
                        & $6.3$     & $0.103\pm0.001$ & $42.776\pm0.053$ & $0.000\pm0.002$ & $(9.224\pm0.006)\times10^{-3}$ & 1/3\\
$K^+$                   & $7.7$     & $0.124\pm0.002$ & $20.001\pm0.051$ & $0.000\pm0.002$ & $(1.262\pm0.020)\times10^{-2}$ & 1/5\\
($u\bar s$)             & $8.8$     & $0.117\pm0.001$ & $24.979\pm0.052$ & $0.000\pm0.001$ & $(1.283\pm0.001)\times10^{-2}$ & 4/7\\
                        & $12.3$    & $0.129\pm0.001$ & $20.018\pm0.051$ & $0.000\pm0.002$ & $(1.443\pm0.012)\times10^{-2}$ & 2/5\\
                        & $17.3$    & $0.132\pm0.001$ & $19.081\pm0.064$ & $0.000\pm0.001$ & $(1.563\pm0.013)\times10^{-2}$ & 1/6\\
\hline
                        & $7.7$     & $0.103\pm0.001$ & $19.411\pm0.051$ & $-0.09\pm0.001$ & $(5.909\pm0.029)\times10^{-4}$ & 12/5\\
$\bar p$                & $8.8$     & $0.101\pm0.001$ & $28.003\pm0.061$ & $-0.09\pm0.003$ & $(8.208\pm0.059)\times10^{-4}$ & 5/1\\
($\bar u\bar u\bar d$)  & $12.3$    & $0.104\pm0.002$ & $25.826\pm0.043$ & $-0.09\pm0.002$ & $(1.714\pm0.014)\times10^{-3}$ & 15/7\\
                        & $17.3$    & $0.128\pm0.001$ & $17.847\pm0.054$ & $-0.09\pm0.002$ & $(3.401\pm0.028)\times10^{-3}$ & 1/16\\
\hline
                        & $6.3$     & $0.103\pm0.001$ & $24.299\pm0.057$ & $-0.09\pm0.002$ & $(3.279\pm0.029)\times10^{-2}$ & 4/5\\
$p$                     & $7.7$     & $0.138\pm0.001$ & $19.765\pm0.056$ & $-0.09\pm0.002$ & $(2.715\pm0.022)\times10^{-2}$ & 0.5/7\\
($uud$)                 & $8.8$     & $0.139\pm0.001$ & $18.627\pm0.044$ & $-0.09\pm0.001$ & $(2.570\pm0.022)\times10^{-2}$ & 1/7\\
                        & $12.3$    & $0.147\pm0.001$ & $12.403\pm0.051$ & $-0.09\pm0.001$ & $(1.583\pm0.013)\times10^{-2}$ & 2/8\\
                        & $17.3$    & $0.147\pm0.001$ & $11.023\pm0.044$ & $-0.09\pm0.001$ & $(1.460\pm0.014)\times10^{-2}$ & 1/6\\
\hline
\end{tabular}%
\end{center}}
\end{table*}

\begin{table*}[!htb]
{\small Table 7. Values of $T$, $n$, $a_{0}$, $N_0$, $\chi^2$, and
ndof corresponding to the curves in Fig. 15. All the free
parameters are extracted at the quark level. \vspace{-0.5cm}
\scriptsize
\begin{center}
\begin{tabular} {cccccccccccc}\\ \hline\hline
Particle & Energy & $T$ (GeV) & $n$ & $a_{0}$ & $N_0$ & $\chi^2$/ndof \\
(quark structure) & (GeV) &  &  &  &  & \\
\hline
                        & $7.7$     & $0.198\pm0.001$ & $31.328\pm0.053$ & $-0.461\pm0.002$ & $24.185\pm0.052$ & 8/22\\
$\pi^-$                 & $11.5$    & $0.197\pm0.001$ & $18.782\pm0.051$ & $-0.461\pm0.002$ & $31.347\pm0.078$ & 3/22\\
($d\bar u$)             & $19.6$    & $0.202\pm0.001$ & $15.002\pm0.053$ & $-0.461\pm0.002$ & $39.916\pm0.052$ & 2/22\\
                        & $27$      & $0.208\pm0.001$ & $15.368\pm0.054$ & $-0.461\pm0.001$ & $43.343\pm0.262$ & 4/22\\
                        & $39$      & $0.216\pm0.001$ & $14.765\pm0.052$ & $-0.461\pm0.001$ & $44.444\pm0.262$ & 1/22\\
\hline
                        & $7.7$     & $0.199\pm0.001$ & $27.972\pm0.047$ & $-0.461\pm0.001$ & $23.299\pm0.131$ & 8/22\\
$\pi^+$                 & $11.5$    & $0.201\pm0.001$ & $18.362\pm0.053$ & $-0.461\pm0.001$ & $29.938\pm0.131$ & 2/22\\
($u\bar d$)             & $19.6$    & $0.204\pm0.001$ & $15.042\pm0.052$ & $-0.461\pm0.001$ & $39.131\pm0.078$ & 1/22\\
                        & $27$      & $0.209\pm0.001$ & $14.512\pm0.061$ & $-0.461\pm0.002$ & $42.245\pm0.262$ & 2/22\\
                        & $39$      & $0.218\pm0.001$ & $13.912\pm0.064$ & $-0.461\pm0.001$ & $43.643\pm0.104$ & 0.4/22\\
\hline
                        & $7.7$     & $0.139\pm0.001$ & $18.572\pm0.053$ & $0.215\pm0.003$ & $1.441\pm0.005$ & 9/19\\
$K^-$                   & $11.5$    & $0.145\pm0.001$ & $16.550\pm0.054$ & $0.215\pm0.001$ & $2.307\pm0.011$ & 6/19\\
($s\bar u$)             & $19.6$    & $0.144\pm0.001$ & $11.102\pm0.063$ & $0.215\pm0.002$ & $3.573\pm0.016$ & 11/22\\
                        & $27$      & $0.149\pm0.001$ & $12.572\pm0.054$ & $0.215\pm0.003$ & $4.314\pm0.018$ & 7/21\\
                        & $39$      & $0.153\pm0.001$ & $10.179\pm0.052$ & $0.215\pm0.001$ & $4.776\pm0.031$ & 6/22\\
\hline
                        & $7.7$     & $0.145\pm0.001$ & $13.392\pm0.057$ & $0.215\pm0.002$ & $3.974\pm0.021$ & 3/19\\
$K^+$                   & $11.5$    & $0.149\pm0.001$ & $13.199\pm0.056$ & $0.215\pm0.002$ & $4.752\pm0.022$ & 6/21\\
($u\bar s$)             & $19.6$    & $0.147\pm0.001$ & $10.660\pm0.062$ & $0.215\pm0.002$ & $5.629\pm0.031$ & 9/22\\
                        & $27$      & $0.152\pm0.001$ & $10.892\pm0.051$ & $0.215\pm0.002$ & $5.942\pm0.031$ & 1/22\\
                        & $39$      & $0.156\pm0.001$ & $11.100\pm0.064$ & $0.215\pm0.001$ & $6.105\pm0.037$ & 4/22\\
\hline
                        & $7.7$     & $0.141\pm0.001$ & $13.393\pm0.063$ & $0.325\pm0.002$ & $0.070\pm0.001$ & 7/11\\
$\bar p$                & $11.5$    & $0.141\pm0.001$ & $13.374\pm0.058$ & $0.325\pm0.002$ & $0.260\pm0.002$ & 9/19\\
($\bar u\bar u\bar d$)  & $19.6$    & $0.147\pm0.001$ & $11.929\pm0.063$ & $0.325\pm0.002$ & $0.754\pm0.006$ & 11/18\\
                        & $27$      & $0.149\pm0.002$ & $11.328\pm0.064$ & $0.325\pm0.003$ & $1.101\pm0.007$ & 15/18\\
                        & $39$      & $0.151\pm0.002$ & $11.028\pm0.052$ & $0.325\pm0.001$ & $1.553\pm0.012$ & 10/19\\
\hline
                        & $7.7$     & $0.142\pm0.001$ & $13.328\pm0.057$ & $0.325\pm0.003$ & $9.542\pm0.093$ & 18/25\\
$p$                     & $11.5$    & $0.141\pm0.001$ & $13.328\pm0.066$ & $0.325\pm0.003$ & $7.838\pm0.066$ & 14/24\\
($uud$)                 & $19.6$    & $0.145\pm0.001$ & $11.919\pm0.062$ & $0.325\pm0.002$ & $6.101\pm0.039$ & 27/25\\
                        & $27$      & $0.147\pm0.002$ & $11.336\pm0.064$ & $0.325\pm0.001$ & $5.506\pm0.041$ & 19/19\\
                        & $39$      & $0.151\pm0.001$ & $10.705\pm0.064$ & $0.325\pm0.003$ & $4.706\pm0.038$ & 17/18\\
\hline
                        & $7.7$     & $0.168\pm0.001$ & $24.852\pm0.055$ & $0.111\pm0.002$ & $12.225\pm0.068$ & 9/8\\
$K^0_{S}$               & $11.5$    & $0.171\pm0.001$ & $20.572\pm0.056$ & $0.111\pm0.002$ & $15.673\pm0.075$ & 12/10\\
($d\bar s$)             & $19.6$    & $0.183\pm0.001$ & $21.191\pm0.062$ & $0.111\pm0.001$ & $20.065\pm0.112$ & 6/11\\
                        & $27$      & $0.189\pm0.001$ & $19.796\pm0.051$ & $0.111\pm0.001$ & $22.010\pm0.075$ & 12/12\\
                        & $39$      & $0.194\pm0.001$ & $17.522\pm0.054$ & $0.111\pm0.001$ & $22.889\pm0.125$ & 6/13\\
\hline
                        & $7.7$     & $0.187\pm0.001$ & $30.125\pm0.063$ & $0.171\pm0.002$ & $0.187\pm0.001$ & 9/6\\
$\bar \Lambda$          & $11.5$    & $0.190\pm0.001$ & $41.964\pm0.066$ & $0.171\pm0.002$ & $0.626\pm0.004$ & 12/9\\
($\bar u\bar d\bar s$)  & $19.6$    & $0.195\pm0.002$ & $33.201\pm0.062$ & $0.171\pm0.002$ & $1.742\pm0.006$ & 29/10\\
                        & $27$      & $0.201\pm0.001$ & $30.202\pm0.064$ & $0.171\pm0.002$ & $2.598\pm0.019$ & 18/11\\
                        & $39$      & $0.205\pm0.001$ & $29.339\pm0.058$ & $0.171\pm0.002$ & $3.609\pm0.025$ & 8/12\\
\hline
                        & $7.7$     & $0.175\pm0.001$ & $37.155\pm0.067$ & $0.171\pm0.002$ & $13.904\pm0.064$ & 4/9\\
$\Lambda$               & $11.5$    & $0.183\pm0.001$ & $34.466\pm0.066$ & $0.171\pm0.002$ & $12.856\pm0.064$ & 8/10\\
($uds$)                 & $19.6$    & $0.194\pm0.001$ & $32.856\pm0.062$ & $0.171\pm0.002$ & $11.357\pm0.070$ & 13/11\\
                        & $27$      & $0.198\pm0.001$ & $30.723\pm0.051$ & $0.171\pm0.003$ & $10.706\pm0.060$ & 19/12\\
                        & $39$      & $0.204\pm0.001$ & $25.445\pm0.054$ & $0.171\pm0.003$ & $2.616\pm0.058$ & 17/12\\
\hline
                        & $7.7$     & $0.162\pm0.001$ & $11.968\pm0.063$ & $0.195\pm0.002$ & $0.064\pm0.004$ & 2/2\\
$\bar \Xi^+$            & $11.5$    & $0.188\pm0.001$ & $23.869\pm0.064$ & $0.195\pm0.003$ & $0.154\pm0.001$ & 2/4\\
($dss$)                 & $19.6$    & $0.196\pm0.001$ & $21.459\pm0.053$ & $0.195\pm0.003$ & $0.385\pm0.003$ & 11/7\\
                        & $27$      & $0.207\pm0.001$ & $28.733\pm0.054$ & $0.195\pm0.002$ & $0.533\pm0.003$ & 8/5\\
                        & $39$      & $0.210\pm0.001$ & $29.019\pm0.062$ & $0.195\pm0.002$ & $0.712\pm0.006$ & 3/5\\
\hline
                        & $7.7$     & $0.158\pm0.001$ & $16.351\pm0.057$ & $0.195\pm0.002$ & $1.071\pm0.009$ & 4/3\\
$\Xi^-$                 & $11.5$    & $0.187\pm0.001$ & $38.872\pm0.055$ & $0.195\pm0.002$ & $1.177\pm0.012$ & 2/4\\
($dss$)                 & $19.6$    & $0.187\pm0.001$ & $21.236\pm0.062$ & $0.195\pm0.001$ & $1.428\pm0.014$ & 3/4\\
                        & $27$      & $0.201\pm0.001$ & $28.249\pm0.061$ & $0.195\pm0.001$ & $1.388\pm0.019$ & 4/5\\
                        & $39$      & $0.211\pm0.001$ & $27.631\pm0.064$ & $0.195\pm0.002$ & $1.354\pm0.012$ & 0.3/5\\
\hline
\end{tabular}%
\end{center}}
\end{table*}

\begin{table*}
{\small Table 8. Values of $T$, $n$, $a_{0}$, $N_0$, $\chi^2$, and
ndof corresponding to the curves in Fig. 16, where ``--" in the
last column denotes ndof = 0 and the corresponding curve is for
the eye guiding only. All the free parameters are extracted at the
quark level. \vspace{-0.5cm}
\begin{center}
\begin{tabular} {cccccccccccc}\\ \hline\hline
Particle & Energy & $T$ (GeV) & $n$ & $a_{0}$ & $N_0$ & $\chi^2$/ndof \\
(quark structure) & (GeV) &  &  &  &  & \\
\hline
                        & $6.3$     & $0.132\pm0.001$ & $4.851\pm0.033$ & $-0.460\pm0.002$ & $14.849\pm0.043$ & 34/12\\
$\pi^-$                 & $7.6$     & $0.148\pm0.001$ & $4.771\pm0.035$ & $-0.460\pm0.001$ & $17.169\pm0.044$ & 28/12\\
($d\bar u$)             & $8.8$     & $0.151\pm0.001$ & $4.771\pm0.036$ & $-0.460\pm0.002$ & $18.662\pm0.074$ & 38/10\\
                        & $12.3$    & $0.151\pm0.001$ & $4.706\pm0.040$ & $-0.460\pm0.002$ & $24.540\pm0.077$ & 16/10\\
                        & $17.3$    & $0.152\pm0.001$ & $4.427\pm0.036$ & $-0.460\pm0.001$ & $31.147\pm0.053$ & 17/10\\
\hline
$\pi^+$                 & $6.3$     & $0.139\pm0.001$ & $5.348\pm0.030$ & $-0.460\pm0.002$ & $13.773\pm0.058$ & 47/12\\
($u\bar d$)             & $7.6$     & $0.152\pm0.001$ & $4.891\pm0.031$ & $-0.460\pm0.001$ & $16.212\pm0.059$ & 34/12\\
\hline
                        & $6.3$     & $0.160\pm0.002$ & $14.142\pm0.043$ & $0.030\pm0.003$ & $0.742\pm0.004$ & 20/6\\
$K^-$                   & $7.6$     & $0.169\pm0.001$ & $12.962\pm0.041$ & $0.030\pm0.001$ & $1.087\pm0.006$ & 62/16\\
($s\bar u$)             & $8.8$     & $0.176\pm0.002$ & $12.962\pm0.038$ & $0.030\pm0.002$ & $1.075\pm0.010$ & 55/6\\
                        & $12.3$    & $0.176\pm0.001$ & $12.962\pm0.039$ & $0.030\pm0.003$ & $1.542\pm0.014$ & 65/14\\
                        & $17.3$    & $0.178\pm0.001$ & $12.999\pm0.032$ & $0.030\pm0.002$ & $2.258\pm0.009$ & 33/8\\
\hline
                        & $6.3$     & $0.177\pm0.001$ & $19.962\pm0.032$ & $0.030\pm0.003$ & $2.270\pm0.013$ & 33/6\\
$K^+$                   & $7.6$     & $0.182\pm0.001$ & $19.002\pm0.038$ & $0.030\pm0.002$ & $2.780\pm0.008$ & 90/6\\
($u\bar s$)             & $8.8$     & $0.182\pm0.001$ & $19.002\pm0.042$ & $0.030\pm0.001$ & $2.956\pm0.013$ & 39/6\\
                        & $12.3$    & $0.182\pm0.001$ & $18.992\pm0.041$ & $0.030\pm0.003$ & $3.390\pm0.007$ & 13/10\\
                        & $17.3$    & $0.187\pm0.001$ & $18.599\pm0.044$ & $0.030\pm0.002$ & $3.822\pm0.027$ & 40/8\\
\hline
$\bar p$                & $8.8$     & $0.220\pm0.002$ & $50.042\pm0.053$ & $0.000\pm0.003$ & $0.095\pm0.002$ & 2/7\\
($\bar u\bar u\bar d$)  & $12.3$    & $0.222\pm0.002$ & $48.023\pm0.051$ & $0.000\pm0.003$ & $0.255\pm0.003$ & 1/7\\
                        & $17.3$    & $0.238\pm0.001$ & $40.923\pm0.053$ & $0.000\pm0.002$ & $0.562\pm0.007$ & 0.3/6\\
\hline
                        & $6.3$     & $0.202\pm0.001$ & $55.040\pm0.045$ & $0.000\pm0.002$ & $16.158\pm0.218$ & 3/11\\
$p$                     & $7.6$     & $0.216\pm0.002$ & $53.015\pm0.046$ & $0.000\pm0.002$ & $13.813\pm0.223$ & 1/11\\
($uud$)                 & $8.8$     & $0.222\pm0.002$ & $53.015\pm0.052$ & $0.000\pm0.002$ & $14.234\pm0.225$ & 2/11\\
                        & $12.3$    & $0.222\pm0.001$ & $53.013\pm0.055$ & $0.000\pm0.002$ & $10.167\pm0.225$ & 2/11\\
                        & $17.3$    & $0.238\pm0.001$ & $43.969\pm0.054$ & $0.000\pm0.001$ & $8.582\pm0.061$ & 1/8\\
\hline
                        & $6.3$     & $0.157\pm0.001$ & $7.992\pm0.046$ & $0.150\pm0.001$ & $0.504\pm0.004$ & 0.4/1\\
$\phi$                  & $7.6$     & $0.179\pm0.001$ & $6.592\pm0.044$ & $0.150\pm0.001$ & $0.504\pm0.003$ & 1/2\\
($s\bar s$)             & $8.8$     & $0.181\pm0.001$ & $6.592\pm0.042$ & $0.150\pm0.001$ & $0.620\pm0.007$ & 11/6\\
                        & $12.3$    & $0.181\pm0.001$ & $6.592\pm0.041$ & $0.150\pm0.002$ & $1.023\pm0.004$ & 6/4\\
                        & $17.3$    & $0.222\pm0.002$ & $6.492\pm0.042$ & $0.150\pm0.002$ & $1.297\pm0.016$ & 28/6\\
\hline
                        & $6.3$     & $0.266\pm0.001$ & $21.640\pm0.041$ & $-0.050\pm0.001$ & $0.031\pm0.001$ & 0.2/1\\
$\bar \Lambda$          & $7.6$     & $0.238\pm0.001$ & $25.643\pm0.048$ & $-0.050\pm0.002$ & $0.085\pm0.001$ & 8/3\\
($\bar u\bar d\bar s)$  & $8.8$     & $0.243\pm0.001$ & $25.502\pm0.053$ & $-0.050\pm0.001$ & $0.136\pm0.002$ & 7/3\\
                        & $12.3$    & $0.235\pm0.001$ & $25.826\pm0.054$ & $-0.050\pm0.001$ & $0.373\pm0.005$ & 7/3\\
                        & $17.3$    & $0.234\pm0.002$ & $25.826\pm0.062$ & $-0.050\pm0.004$ & $0.407\pm0.011$ & 40/5\\
\hline
                        & $6.3$     & $0.198\pm0.001$ & $29.960\pm0.038$ & $-0.050\pm0.001$ & $5.570\pm0.010$ & 22/12\\
$\Lambda$               & $7.6$     & $0.200\pm0.001$ & $27.999\pm0.056$ & $-0.050\pm0.001$ & $6.531\pm0.077$ & 47/12\\
($uds$)                 & $8.8$     & $0.210\pm0.001$ & $27.999\pm0.052$ & $-0.050\pm0.003$ & $6.165\pm0.052$ & 54/12\\
                        & $12.3$    & $0.222\pm0.002$ & $27.391\pm0.071$ & $-0.050\pm0.004$ & $5.748\pm0.079$ & 128/11\\
                        & $17.3$    & $0.232\pm0.001$ & $27.391\pm0.064$ & $-0.050\pm0.005$ & $4.354\pm0.053$ & 77/13\\
\hline
                        & $6.3$     & $0.185\pm0.001$ & $27.899\pm0.057$ & $-0.050\pm0.002$ & $0.432\pm0.006$ & 14/2\\
$\Xi^-$                 & $7.6$     & $0.187\pm0.002$ & $26.364\pm0.063$ & $-0.050\pm0.004$ & $0.523\pm0.009$ & 4/2\\
($dss$)                 & $8.8$     & $0.189\pm0.002$ & $25.886\pm0.062$ & $-0.050\pm0.004$ & $0.521\pm0.006$ & 12/3\\
                        & $12.3$    & $0.203\pm0.001$ & $20.774\pm0.055$ & $-0.050\pm0.003$ & $0.608\pm0.006$ & 26/2\\
                        & $17.3$    & $0.204\pm0.002$ & $18.744\pm0.064$ & $-0.050\pm0.003$ & $0.661\pm0.006$ & 5/4\\
\hline
                        & $7.6$     & $0.188\pm0.002$ & $26.364\pm0.061$ & $-0.050\pm0.004$ & $0.025\pm0.001$ & 1/--\\
$\bar \Xi^+$            & $8.8$     & $0.196\pm0.002$ & $20.062\pm0.058$ & $-0.050\pm0.004$ & $0.033\pm0.001$ & 2/--\\
($dss$)                 & $12.3$    & $0.196\pm0.001$ & $20.064\pm0.063$ & $-0.050\pm0.002$ & $0.093\pm0.001$ & 0.2/1\\
                        & $17.3$    & $0.225\pm0.001$ & $17.012\pm0.064$ & $-0.050\pm0.002$ & $0.157\pm0.002$ & 5/2\\
\hline
\end{tabular}%
\end{center}}
\end{table*}
\begin{multicols}{2}

The transverse mass ($m_{\rm T}=\sqrt{p_{\rm T}^2+m_0^2}$) (minus
rest mass $m_0$) spectra, $(1/m_{\rm T})d^2N/dm_{\rm T}dy$, of (a)
$\pi^+$ and $\pi^-$, (b) $K^+$ and $K^-$, (c) $p$ and $\bar p$,
(d) $\phi$, (e) $\Lambda$ and $\bar\Lambda$, as well as (f)
$\bar\Xi^+$ and $\Xi^-$ with (a)--(c), (e), and (f) $0<y<0.2$, as
well as (d) $0<y<1.0$ to $0<y<1.8$ produced in central (0--5\% to
0--10\%) Pb-Pb collisions at 6.3, 7.6, 8.8, 12.3, and 17.3 GeV are
shown in Fig. 16. The symbols represent the experimental data
measured by the NA49 Collaborations~\cite{72,72a,72b,72c,72d} at
the SPS and the curves are our fitted results by using Eqs. (3)
for mesons and (5) for baryons. The related parameters are listed
in Table 8 and the curves fit well the experimental data measured
by the NA49 Collaboration.

Comparing with high energy collisions, low energy collisions show
low effective temperature $T$. Over an energy range decreasing
from the LHC to SPS or RHIC BES, the entropy index-related $n$
increases and the revised index $a_0$ do not show significant
variation. The conclusions obtained from high energy collisions
are coordinated with those from low energy collisions, though the
energy dependent $T$ and $n$ are observed. It is natural that low
energy collisions result in low effective temperature due to low
energy deposition during the collision process. At low energy, the
long collision time renders closer to equilibrium, in which $n$ is
large and $q$ is small.

The collisions at SPS or RHIC BES provide an opportunity to study
the dependence of $T$ and $n$ on $\sqrt{s_{\rm NN}}$ around the
critical point/region of the deconfinement phase transition from
hadronic matter to QGP. Our exploratory test shows that two
parameterized functions are needed for the relation of $T$ versus
$\sqrt{s_{\rm NN}}$. In the region of $\sqrt{s_{\rm NN}}<8$ GeV,
$T=a_1[1-\exp(-\sqrt{s_{\rm NN}}/b_1)]$. In the region of
$\sqrt{s_{\rm NN}}\geq8$ GeV, $T=a_2\ln(\sqrt{s_{\rm NN}})+b_2$.
Here, $T$, $\sqrt{s_{NN}}$, and four parameters ($a_{1,2}$ and
$b_{1,2}$) are in the units of GeV. The parameters are particle
dependent due to individual fitting. We have not used the
simultaneous fitting in the present work due to unsatisfactory
results.

Meanwhile, single parameterized function, $n=a_3\exp(-\sqrt{s_{\rm
NN}}/b_3)+c_3$, is needed for the relation of $n$ versus
$\sqrt{s_{\rm NN}}$, where the parameters are particle dependent,
$a_3$ and $c_3$ are dimensionless, and $b_3$ is in the units of
GeV. The increase of $n$ means the decrease of $q$ at lower
energy, where the system is closer to the equilibrium due to
enough and longer interaction time.

Because very limited collision systems at given energy are
studied, the parameterized functions of $T$ and $n$ versus system
size are not available here. Generally, with the increase of
system size, $T$ and $n$ increase slightly or do not show
significant change. For large system, more energies were deposited
and more multiple scatterings had happened in the collisions,
which cases $T$ and $n$ to have tendency of slight increase.
\\

\end{multicols}
\end{document}